


 \tolerance=10000
  \input phyzzx.tex
\def\W {{\cal W }}

 \Pubnum = {QMW/PH/92/1}
 \date = {January 1992}
 \pubtype={}
\titlepage
\title {\bf CLASSICAL AND QUANTUM \W -GRAVITY}

\author {C. M. Hull}
\
\address {Physics Department,
Queen Mary and Westfield College,
\break
Mile End Road, London E1 4NS, United Kingdom.}

\abstract
{Classical \W-gravities and the corresponding quantum theories
are reviewed. \W-gravities are higher-spin
gauge theories in two dimensions whose gauge algebras are \W-algebras.
 The  geometrical structure of classical \W-gravity
 is investigated, leading to surprising
 connections with self-dual geometry. The anomalies that arise in quantum
\W-gravity
 are discussed, with particular attention to the new
 types of anomalies that arise for non-linearly realised symmetries
 and to the relation between path-integral anomalies and
 non-closure of the quantum current algebra. Models in which
 all anomalies are cancelled by ghost contributions lead to
 new generalisations of string theories.}

  \endpage

 \pagenumber=1


 \REF\pol{A.M. Polyakov,   Phys. Lett. {\bf 103B} (1981) 207,211.}
\REF\he{C.M. Hull, \pl\ {\bf 240B} (1990) 110.}
 \REF\van{K. Schoutens, A. Sevrin and
P. van Nieuwenhuizen, \pl\ {\bf 243B} (1990) 245.}
\REF\hee{C.M. Hull,  \np\ {\bf 353
B} (1991) 707.}
\REF\heee{C.M. Hull, \pl\ {\bf 259B} (1991) 68.}
\REF\vann{K. Schoutens, A.
Sevrin and P. van Nieuwenhuizen, Nucl.Phys. {\bf B349}
(1991) 791 and Phys.Lett. {\bf 251B} (1990) 355.}
\REF\heeee{C.M. Hull, \np\ {\bf 364
B} (1991) 621.}
 \REF\pop{E. Bergshoeff, C.N. Pope, L.J. Romans, E.
Sezgin, X. Shen and K.S. Stelle, \pl\ {\bf 243B}
(1990) 350}
\REF\popp{ E. Bergshoeff, C.N. Pope and K.S. Stelle, \pl\ {\bf
249B} (1990) 208.}
\REF\miko {A. Mikovi\' c, \pl\ {\bf 260B} (1991) 75.}
\REF\mikoham {A. Mikovi\' c,
QMW preprint, QMW/PH/91/13 (1991).}
\REF\hegeom{C.M. Hull, \pl\ {\bf 269B} (1991) 257.}
\REF\zam{A.B. Zamolodchikov,
Teor. Mat. Fiz. {\bf 65} (1985) 1205.}
\REF\fat{V.A. Fateev and A.B.
Zamolodchikov, \np\ {\bf B280} (1987) 644.}
\REF\fatty{V.A. Fateev and S.
Lykanov, \intmp\ {\bf A3} (1988) 507.}
\REF\bil{A. Bilal and J.-L. Gervais, \pl\ {\bf
206B} (1988) 412; \np\ {\bf B314} (1989) 646; \np\ {\bf B318} (1989) 579;
Ecole Normale preprint LPTENS 88/34.}
\REF \li{K. Li, Caltech preprint  CALT-68-1724 (1991).}
\REF\bou{P. Bouwknegt, \pl\  {\bf 207B} (1988) 295.}
\REF\ham{K. Hamada and M. Takao, \pl\ {\bf 209B} (1988) 247; Erratum \pl\
{\bf 213B} (1988) 564.}
 \REF\ofar{J.M. Figueroa-O'Farrill and S. Schrans, \pl\ {\bf 245B} (1990) 471.}
\REF\winf{I. Bakas, \pl\ {\bf 228B} (1989) 57 and Maryland preprint
MdDP-PP-90-033.}
\REF\qwinf{C.N. Pope, L.J. Romans and X. Shen, \pl\ {\bf 236B}
(1990) 173; \np\ {\bf B339} (1990) 191.}
\REF\wittt{E. Witten, \cmp\ {\bf 92} (1984) 455.}
\REF\bais{F. Bais, P. Bouwknegt, M. Surridge and K. Schoutens, \np\ {\bf
B304} (1988) 348, 371.}
\REF\thierrya{J. Thierry-Mieg, Lectures given at the Cargese School on
Non-Perturbative QFT, (1987).}
\REF\gun{M. G\" unaydin, G. Sierra and P.K. Townsend,
\np\ {\bf B242} (1984)
244 and  \np\ {\bf B253} (1985) 573.}
\REF\romans{L.J. Romans, \np\ {\bf B352} 829.}
\REF\schafer{R.D. Schafer, J. Math. Mech. {\bf 10} (1) (1969) 159.}
 \REF\howpapa{P. Howe and G. Papadopolous, \pl\ {\bf B267} (1991) 362,
 \pl\ {\bf B263} (1991) 230 and QMW preprints.}
\REF\basti{Bastianelli, \mpl\ {\bf A6} (1991) 425.}
\REF\wit{E. Witten, IAS preprint IASSNS-HEP-90/37 (1990).}
\REF\wot{A. Bilal, \pl\ {\bf 249B} (1990) 56;
A. Bilal, V.V. Fock and I.I. Kogan, \np\ {\bf B359} (1991) 635.}
\REF\sot{G. Sotkov and M. Stanishkov, \np\ {\bf B356} (1991) 439;
G. Sotkov, M. Stanishkov and C.J. Zhu, \np\ {\bf B356} (1991) 245.}
\REF\riemy{B. Riemann, (1892) \lq Uber die Hypothesen welche der Geometrie
zugrunde liegen', Ges. Math. Werke, Leibzig.}
\REF\finre{H. Rund, \lq The Differential Geometry of Finsler Spaces', Nauka,
Moscow (1981); G.S. Asanov, \lq Finsler Geometry, Relativity and Gauge
Theories', Reidel, Dordrecht (1985).}
\REF\pen{R. Penrose, Gen. Rel. and Grav., {\bf 7} (1976) 31.}
\REF\hit{U. Lindstrom and M. Ro\v cek, \np\ {\bf B222} (1983) 285;
N. Hitchin, A. Karlhede, U. Lindstrom and M. Ro\v cek, Comm. Math. Phys.
{\bf 108} (1987) 535. }
 \REF\park{Q-Han Park, \pl\ {\bf 236B} (1990) 429, {\bf 238B}
(1990) 287 and {\bf 257B} (1991) 105; K. Yagamishi and G.F. Chapline,
Class. Quantum Grav., {\bf 8} (1991) 427 and
\pl\ {\bf 259B} (1991) 463.}
\REF\kir{I. Bakas and E. Kiritsis, \np\ {\bf B343} (1990) 185.}
\REF\thierry{J. Thierry-Mieg, \pl\ {\bf 197B} (1987) 368.}
\REF\vanbrs{K. Schoutens, A.
Sevrin and P. van Nieuwenhuizen, \cmp\ {\bf 124} (1989) 87.}
\REF\wanog{C.M. Hull, \np\ {\bf B367} (1991) 731.}
\REF\wama{C.M. Hull, \pl\ {\bf 259B} (1991) 68.}
\REF\awa{M. Awada and Z. Qiu, \pl\ {\bf 245B} (1990) 85 and Florida preprints
(1990);
  M. Berschadsky and H. Ooguri,
\cmp\ {\bf 126} (1989) 49;
P. Mansfield and B. Spence, ITP preprint (1990) NSF-ITP-90-242;
J. Pawelczyk, \pl\ {\bf 255B} (1991) 330.}
\REF\mat{Y. Matsuo, \pl\ {\bf 227B} (1989) 209.}
\REF\mato{Y. Matsuo, in the proceedings of the meeting \lq Geometry and
Physics', Lake Tahoe  (1989).}
\REF\wstrbil{A. Bilal and J.-L. Gervais, \np\ {\bf B326}
(1989) 222.}
\REF\lipop{K. Li and C.N. Pope, Class. Quantum Grav. {\bf 8} (1991) 1677.}
\REF\zet{K. Yagamishi, \pl\ {\bf  266B}  (1991) 370;
C.N. Pope, L.J. Romans and X. Shen, \pl\ {\bf  254B}  (1991) 401.}
\REF\gates{S.J. Gates, P.S. Howe and C.M. Hull,
\pl\ {\bf 227B} (1989) 49.}
 \REF\vanan{K. Schoutens, A.
Sevrin and P. van Nieuwenhuizen, \np\ {\bf B364} (1991) 584.}
\REF\vanangh{K. Schoutens, A.
Sevrin and P. van Nieuwenhuizen, in Proceedings of
1991 Miami Workshop, (Plenum, New York, 1991).}
\REF\pal{C. M. Hull and L. Palacios,
   Mod. Phys. Lett. {\bf 6} (1991) 2995.}
\REF\hegho{C.M. Hull, QMW preprint QMW/PH/91/14 (1991), to appear
in \intmp.}
\REF\vanangha{K. Schoutens, A.
Sevrin and P. van Nieuwenhuizen, Stony Brook preprints
ITP-SB-91-21, ITP-SB-91-50;
H. Ooguri, K. Schoutens, A.
Sevrin and P. van Nieuwenhuizen, Stony Brook preprint
ITP-SB-91-16.}
\REF\deform{E. Bergshoeff, P.S. Howe, C.N. Pope,  E.
Sezgin, X. Shen and K.S. Stelle,
Nucl. Phys. {\bf B363}, (1991) 163.}
\REF\frau{A. Ceresole, M. Frau and A. Lerda, Phys. Lett.
{\bf 265B}, (1991) 72. }
\REF\das{S.R. Das, A. Dhar and S.K. Rama, Tata preprints (1991)
TIFR/TH/91-11,TIFR/TH/91-20.}
\REF\wstr{C.N. Pope, L.J. Romans and K.S. Stelle, \pl\ {\bf 268B}
(1991) 167 and {\bf 269B}
(1991) 287.}
\REF\wstrspec{C.N. Pope, L.J. Romans and K.S. Stelle, Texas A \& M preprint,
CTP TAMU-68/91.}
\REF\chirgag{A.M. Polyakov, Mod. Phys. Lett. {\bf A2} (1987) 443.}
\REF\batalina{I.A. Batalin and G.A. Vilkovisky, \pl\ {\bf 102B} (1981) 27
and \pr\ {\bf D28} (1983) 2567.}
\REF\brink{P.S. Howe, U. Lindstrom and P. White, \pl\ {\bf 246B} (1990) 430.}
\REF\baswar{F. Bastianelli, \pl\ { \bf 263B} (1991) 411.}

\def\sqr#1#2{{\vcenter{\hrule height.#2pt
      \hbox{\vrule width.#2pt height#1pt \kern#1pt
        \vrule width.#2pt}
      \hrule height.#2pt}}}

 \def\gtorder{\mathrel{\raise.3ex\hbox{$>$}\mkern-14mu
             \lower0.6ex\hbox{$\sim$}}}
\def\ltorder{\mathrel{\raise.3ex\hbox{$<$}|mkern-14mu
             \lower0.6ex\hbox{\sim$}}}
\def\dalemb#1#2{{\vbox{\hrule height .#2pt
        \hbox{\vrule width.#2pt height#1pt \kern#1pt
                \vrule width.#2pt}
        \hrule height.#2pt}}}
\def\square{\mathord{\dalemb{5.9}{6}\hbox{\hskip1pt}}}
\def\boxx{\square}

\def\np{Nucl. Phys.}
\def\pl{Phys. Lett.}
\def\pr{Phys. Rev.}

\def\cmp{Comm. Math. Phys.}
\def\intmp{Intern. J. Mod. Phys.}
\def\mpl{Mod. Phys. Lett.}

\def\half{{\textstyle {1 \over 2}}}

\def\dm {\partial_{\mu}}
\def\dn {\partial_{\nu}}
\def\dr {\partial_{\rho}}

\def\ffi {\phi^i}
\def\fj {\phi^j}
\def\fk {\phi^k}
\def\ix {\int\!\!d^2x\;}
\def\intt {\int\!\! }

\def\dpl {\partial_+}
\def\dmi {\partial_-}

\def\lie  { {\cal L}  }

\def\nox{{\scriptstyle{\times \atop \times}}}
\def\op{\nox}
\def\cl{\nox}

\def\IB{\relax{\rm I\kern-.18em B}}
\def\IC{\relax\hbox{$\inbar\kern-.3em{\rm C}$}}
\def\ID{\relax{\rm I\kern-.18em D}}
\def\IE{\relax{\rm I\kern-.18em E}}
\def\IF{\relax{\rm I\kern-.18em F}}
\def\IG{\relax\hbox{$\inbar\kern-.3em{\rm G}$}}
\def\IH{\relax{\rm I\kern-.18em H}}
\def\II{\relax{\rm I\kern-.18em I}}
\def\IK{\relax{\rm I\kern-.18em K}}
\def\IL{\relax{\rm I\kern-.18em L}}
\def\IM{\relax{\rm I\kern-.18em M}}
\def\IN{\relax{\rm I\kern-.18em N}}
\def\IO{\relax\hbox{$\inbar\kern-.3em{\rm O}$}}
\def\IP{\relax{\rm I\kern-.18em P}}
\def\IQ{\relax\hbox{$\inbar\kern-.3em{\rm Q}$}}
\def\IR{\relax{\rm I\kern-.18em R}}

\chapter{Introduction}

There
has been considerable interest in the infinite-dimensional algebras
that can arise as symmetries of two-dimensional field theories
and there is an intimate relationship between such algebras  and
two-dimensional
gauge theories or string theories.
Perhaps the most important example is the Virasoro algebra, which is a
symmetry algebra of any two dimensional conformal field theory.
This infinite-dimensional rigid symmetry can be promoted to a local symmetry
(two-dimensional diffeomorphisms) by coupling the two-dimensional field
theory to gravity, resulting in a theory that is Weyl-invariant as well as
diffeomorphism invariant. The two-dimensional metric  enters the theory as
a Lagrange multiplier imposing   constraints which satisfy the Virasoro
algebra, and the Virasoro algebra also emerges as the residual symmetry
that remains after choosing a conformal gauge.
 The quantisation of such a system of matter coupled to gravity
defines a   string theory and if the matter system is chosen such that
the world-sheet metric $h_{\mu \nu}$ decouples from the quantum theory
(\ie\ if the matter central charge is $c=26$), the string theory is said to
be critical [\pol]. Remarkably, the introduction of   gravity on the
world-sheet
leads to a critical string theory which leads to gravity in space-time.

The situation is similar for each of the cases in the following table.
\medskip
\smallskip
{\Tenpoint {
\def\thalf{{\textstyle {3 \over 2}}}
\settabs\+ &   $N=2$ Super-Virasoro ~   &  $2,\thalf,\thalf,1$   ~ &
  Topological Gravity  ~ &   $h_{\mu
\nu},\psi_\mu, \bar \psi_\mu, A_\mu$  ~ & \it $N=2$ Superstring
  \cr \+&\it Algebra &\it Spins  &\it 2-D Gauge Theory   &\it
Gauge Fields  & \it String Theory \cr
\smallskip
\+&Virasoro Algebra &2& Gravity &$h_{\mu \nu}$ &Bosonic String\cr
\+&Super-Virasoro  &$2,\thalf$& Supergravity &$h_{\mu \nu},\psi_\mu$
&Superstring\cr
\+&$N=2$ Super-Virasoro  &$2,\thalf,\thalf,1$&$N=2$ Supergravity &$h_{\mu
\nu},\psi_\mu, \bar \psi_\mu, A_\mu$ &$N=2$ Superstring\cr
\+&Topological Virasoro   &$2,2,1,1$& Topological Gravity &$h_{\mu
\nu},g_{\mu
\nu}, A_\mu,\psi_\mu$ &Topological String\cr
\+&\W-Algebra   &$2,3,\dots$&\W-Gravity &$h_{\mu
\nu},B_{\mu
\nu \rho}, \dots$ &\W-Strings\cr  }}
\smallskip
\medskip
\noindent In the first column are extended conformal algebras, \ie\ infinite
dimensional algebras that contain the Virasoro algebra. Each algebra is
generated by a set of currents, whose spins are labelled in the second
column. Each algebra can arise as the symmetry algebra of a particular class
of conformal field theories e.g. the super-Virasoro algebra is a symmetry of
super-conformal field theories
while the topological Virasoro algebra is a
symmetry of topological
conformal field theories. For such
theories, the infinite-dimensional rigid
symmetry of the matter system
can be promoted to a local symmetry by coupling to the gauge theory
listed in the third column. In this coupling, the currents generating the
extended conformal algebra are coupled to the  corresponding gauge fields in
the fourth column. In each case, the gauge fields enter as Lagrange
multipliers and the constraints that they impose satisfy the algebra given in
the first column. Finally, integration over the matter and gauge fields
defines a generalisation of string theory which is listed in the last column.
In general, the gauge fields will become dynamical in the quantum theory, but
for special choices of conformal matter system (e.g. $c=26$ systems for the
bosonic string, $c=0$ systems for the topological string or $c=15$ systems
for the $N=1$ superstring), the string theory will be \lq critical' and the
gauge fields will decouple from the theory. A row can be added to the table
corresponding to any two-dimensional extended conformal algebra.

The subject of this review is the set of models corresponding to the last row
of
the table. A \W-algebra might be defined as any extended conformal algebra,
\ie\ a closed algebra that satisfies the Jacobi identities,   contains
the Virasoro algebra as a subalgebra and is generated by a (possibly
infinite) set of chiral currents. Often the definition of \W-algebra
is restricted    to those algebras for which at least one of the generating
currents has spin greater than $2$, but relaxing this condition allows the
definition to include all the algebras in the  table and almost all of the
results to be reviewed here apply with this more general definition.
However,
many (but not all) interesting \W-algebras contain a spin-three
current, and for this reason $3$ is included as a typical higher spin
in the \W-algebra entry in the
  table.

The simplest \W-algebras are those that are Lie algebras, with the generators
$t_a$ (labelled by an index $a$ which will  in general have an infinite range)
satisfying commutation relations of the form
$$
[t_a,t_b]={f_{ab}}^ct_c +c_{ab}
\eqn\lie$$
for some structure constants ${f_{ab}}^c$ and constants $c_{ab}$, which define
a central extension of the algebra. However, for many \W-algebras, the
commutation relations give a result non-linear in the generators
$$
[t_a,t_b]={f_{ab}}^ct_c +c_{ab}+{g_{ab}}^{cd}t_ct_d+\dots =F_{ab}(t_c)
\eqn\nonlie$$
and the algebra can be said to close in the sense that the right-hand-side is a
function of the generators.
Most of the \W-algebras that are generated by a finite number of currents,
with at least one current of spin greater than two, are non-linear algebras of
this type.
Classical \W-algebras for which the bracket in \nonlie\ is a Poisson
bracket are straightforward to define, as the non-linear terms on the
right-hand-side can be taken to be a   product of classical
charges. For quantum \W-algebras, however,   the bracket is realised as a
commutator of quantum operators and the definition of the right-hand-side
requires some normal-ordering prescription. The complications associated
with the normal-ordering mean that there are classical \W-algebras for which
there is no corresponding quantum \W-algebra that satisfies the Jacobi
identities [\hee]. At first sight, it appears that there might be a problem in
trying to realise a non-linear algebra in a field theory, as symmetry
algebras are usually Lie algebras. However, as will be seen, a non-linear
algebra can be realised as a symmetry algebra for which the structure \lq
constants' are replaced by field-dependent quantities.

Consider a field theory in flat Minkowski space with metric $\eta_{\mu \nu}$
and coordinates $x^0,x^1$. The stress-energy tensor is a symmetric tensor
$T_{\mu \nu}$ which, for a translation-invariant theory, satisfies the
conservation law $$
\partial ^\mu
T_{\mu \nu}=0
\eqn\tcon$$
A spin-$s$ current in flat two-dimensionsal space is a rank-$s$ symmetric
tensor $W_{\mu_1 \mu_2 \dots \mu _s}$\foot{Recall that, in two dimensions, any
tensor can be decomposed into a set of symmetric tensors, e.g.
$V_{\mu \nu}=V_{(\mu \nu)}+V\epsilon _{\mu \nu}$ where $V=\half \epsilon ^{\mu
\nu}V_{\mu \nu}$. Thus without loss of generality, all the conserved
currents of a given theory can be taken to be symmetric tensors. A rank-$s$
symmetric tensor transforms as the spin-$s$ representation of the
two-dimensional Lorentz group.} and will be conserved if $$ \partial ^{\mu
_1} W_{\mu_1 \mu_2 \dots \mu _s}=0 \eqn\wcon$$ A theory is conformally
invariant if the stress tensor is traceless, ${T_\mu}^\mu=0$. Introducing
null coordinates $x^\pm={1 \over \sqrt 2} (x^0 \pm x^1)$, the tracelessness
condition becomes $T_{+-}=0$ and \tcon\ then implies that the remaining
components $T_{\pm\pm}$ satisfy
$$ \dpl T_{--}=0, \qquad \dmi T_{++}=0
\eqn \rert$$
If a spin-$s$ current $W_{\mu_1 \mu_2 \dots \mu _s}$ is traceless, it will
have only two non-vanishing components, $W_{++ \dots  +}$
and $W_{--\dots -}$. The conservation condition \wcon\ then
implies that
$$ \dmi W_{++ \dots  +}=0, \qquad \dpl W_{--\dots -}=0\eqn\dfg$$
so that $W_{++ \dots  +}=W_{++ \dots  +}(x^+)$
and $W_{--\dots -}=W_{--\dots -}(x^-)$ are right- and left-moving chiral
currents, respectively.
For a given conformal field theory, the set of all right-moving chiral
currents generate a closed current algebra, the right-moving chiral algebra,
and similarly for left-movers. The right and left chiral algebras are
examples of \W-algebras but are often too large to be useful. In studying
conformal field theories, it is often useful to restrict attention to all
theories whose chiral algebras contain a particular \W-algebra; the
representation theory of that \W-algebra then gives a great deal of useful
information about the spectrum, modular invariants etc of those theories
and may lead to a classification.

A field theory with action $S_0$ and symmetric tensor conserved currents
$T_{\mu \nu},W^A_{\mu_1 \mu_2 \dots \mu _{s_A}}$
(where $A=1,2,\dots $ labels the currents, which have spin $s_A$) will be
invariant under rigid symmetries  with constant parameters $k^\mu, \lambda
_A
^{\mu_1 \mu_2 \dots \mu _{s_A-1}}$ (translations and \lq
\W-translations') generated by the Noether charges $P_\mu,
 Q^A_{\mu_1 \mu_2 \dots \mu _{s_A-1}}$
(momentum and \lq \W-momentum') given by
  $P_\mu=\intt
dx^0T_{0\mu}$ and  $ Q^A_{\mu_1 \mu_2 \dots \mu _{s_A-1}}=\intt dx^0
W^A_{\mu_1 \mu_2 \dots \mu _{s_A-1}0}$. This is
true of non-conformal theories (e.g. affine Toda theories) as well as
conformal ones. However, if   the currents are traceless, then the theory
is in fact invariant under an infinite dimensional extended conformal
symmetry. The parameters $ \lambda
_A
^{\mu_1 \mu_2 \dots \mu _{s_A-1}}$
are then
 traceless and the corresponding transformations will be symmetries if the
parameters are not constant but satisfy the conditions that the
trace-free parts of $\partial ^{(\nu}k^{\mu)}, \partial ^{(\nu}
 \lambda
_A
^{\mu_1 \mu_2 \dots \mu _{s_A-1})}$ are zero.
This implies that
$\partial_{\mp}k^ \pm=0$ and $ \partial_{\mp}
\lambda _A^{\pm \pm \dots \pm} = 0$ so that
 the parameters are \lq semi-local', $k^\pm = k^\pm(x^\pm)$ and $\lambda
_A^{\pm \pm \dots \pm} =  \lambda _A^{\pm \pm \dots \pm}(x^\pm)$
and these are the parameters of conformal and \lq \W-conformal'
transformations.

The rigid symmetries corresponding to the currents $T_{\mu \nu},W^A_{\mu_1
\mu_2 \dots \mu _{s_A}}$ can be promoted to local ones by coupling to
the \W-gravity gauge fields
$h^{\mu \nu}, B_A^{\mu_1
\mu_2 \dots \mu _{s_A}}$ which are symmetric tensors transforming as
$$
\delta h^{\mu \nu}=\partial ^{(\nu}k^{\mu)}+\dots ,\qquad
\delta B_A^{\mu_1
\mu_2 \dots \mu _s}=\partial ^{(\nu}
\lambda_A ^{\mu_1 \mu_2 \dots \mu _{s-1})}+\dots,
\eqn\traaaa$$
to lowest order in the gauge fields. The action is given by the Noether
coupling
$$
S=S_0+ \ix \left( h^{\mu \nu}T_{\mu \nu}+ B_A^{\mu_1
\mu_2 \dots \mu _{s_A}}W^A_{\mu_1
\mu_2 \dots \mu _{s_A}}\right) + \dots
\eqn\snoth$$
plus terms non-linear in the gauge fields. If the currents
$T_{\mu \nu},W^A_{\mu_1
\mu_2 \dots \mu _s}$
are traceless, \ie\ if there is extended conformal symmetry, then the traces of
the gauge fields decouple and the theory is invariant under Weyl and \lq
\W-Weyl' transformations given to lowest order in the gauge fields by
$$
\delta h^{\mu \nu}= \Omega \eta  ^{\mu \nu}+\dots ,
\qquad
\delta B_A^{\mu_1
\mu_2 \dots \mu _s}=\Omega_A ^{(\mu_1
\mu_2 \dots \mu _{s-2}}\eta  ^{\mu _{s-1}\mu_s)}+\dots
\eqn\wweyl$$
where $\Omega(x^\nu)$, $\Omega_A^{ \mu_1
\mu_2 \dots \mu _{s-2}}(x^\nu)$ are the local parameters.
This defines the linearised coupling to \W-gravity. The full non-linear
theory is in general non-polynomial in the gauge fields of spins 2 and higher.
The non-linear theory can be constructed to any given order using the Noether
method, but to obtain the full non-linear structure requires a deeper
understanding of the geometry underlying \W-gravity.

This review falls into three main parts. In the first part (chapters 2-5)
classical \W-algebras and \W-gravity will be discussed. In the second part,
(chapter 6), a geometric approach will be used to obtain the full
non-linear structure  of a particular \W-gravity theory.
In the final part, (chapters 7-10) the quantisation of \W-gravity coupled to
\W-matter will be investigated and the anomaly structure described. These
results will then be used to consider the construction of string theories
based on \W-algebras.

\chapter{Classical \W-Algebras}

Consider a set {\cal S} of classical right-moving chiral currents
$T(x^+)=T_{++}(x^+),W(x^+),\dots$ of spins $2,s_W,\dots$. As the currents are
classical, there is no problem in defining products of currents. Suppose there
is also a (graded) anti-symmetric bracket product $[A,B]$ defined on the set
of currents. The main example that will be of interest here is that in which
the currents arise from some classical  field theory and the bracket is the
Poisson bracket in a canonical formalism in which $x^-$ is regarded as the
time variable. The current $T$ satisfies the conformal algebra if
$$ [T(x^+),T(y^+)]= -
\delta ^ \prime
(x^+-y^+)[T(x^+)+T(y^+)]+{c \over 12}
\delta^{ \prime\prime\prime}(x^+-y^+)\eqn\con$$
in which case its modes $L_n$ generate the Virasoro algebra with central
charge $c$.
A current $W$ is said to be primary of spin $s_W$ if
$$ [T(x^+),W(y^+)]= -
\delta ^ \prime
(x^+-y^+)[W(x^+)+(s_W-1)W(y^+)]
\eqn\prim$$
The set {\cal S} of currents will generate a \W-algebra if the bracket of any
two currents gives a function of currents in {\cal S} and if the bracket
satisfies the Jacobi identities.

Consider first the  case in which there are just two currents, $T$
and $W$, where $W$ is  primary of spin $s=s_W$.
For simplicity, suppose that $c=0$ (at least classically) and that the
$[W,W]$ bracket takes the form
$$ [W(x^+),W(y^+)]= -2
\kappa \delta ^ \prime
(x^+-y^+)[\Lambda(x^+)+\Lambda(y^+)]
\eqn\walg$$
 for some $\Lambda$, where $\kappa$ is a constant. If the algebra is to
close, the current $\Lambda$ must be a function of the currents $T,W$ and
their derivatives. If $s=3$, then $\Lambda$ is a spin-four current and the
Jacobi identities are satisfied if
$$
\Lambda=TT\eqn\lamis$$
The algebra then closes non-linearly to give a certain classical limit of the
$\W_3$ algebra of Zamolodchikov [\zam]. (In the limit $\kappa \rightarrow 0$,
this contracts to a linear algebra [\hee, \li].)

For $s>3$,  the algebra will again close and satisfy the Jacobi identities
if $\Lambda $  depends on $T,W$ but not on their derivatives.
If
$s$ is even, the most general such $\Lambda $
is  of the form
$$
\Lambda = \alpha T^{s-1} +\beta WT^{s/2-1}
\eqn\lop$$
for some constants $\alpha ,\beta$,
while if $s$ is  odd, such a $\Lambda $ must
be of the form \lop\ with $\beta=0$.
The   algebra given by \con,\prim,\walg\ and \lop\ is the algebra $\W_{s/s-2}$
of ref. [\hee].  For $s=3$, it
is a classical limit of the $\W _3$ algebra while
for $s=4$ and $s=6$ it
 is a classical limit of the
quantum algebras conjectured to exist by Bouwknegt
[\bou] and constructed in
[\ham,\ofar].
 For all other integer values of $s(>4)$, the argument of
Bouwknegt [\bou] shows that there can be no  quantum operator algebra   that
satisfies the Jacobi identities for generic values of the Virasoro central
charge and for which
\con,\prim,\walg,\lop\ is a classical limit.

A large number of \W-algebras are now known. The $\W_N$ algebra [\fatty,\bil]
 has
currents of spins $2,3,\dots ,N$ (so that $\W_2$ is the Virasoro algebra), the
$\W_\infty $ [\winf,\qwinf]
algebra has currents of spins $2,3,\dots ,\infty$ while the
$\W_{1+\infty} $ algebra [\qwinf] has
currents of spins $1,2,3,\dots ,\infty$.
Each of these algebras
  is a classical limit of a quantum
algebra. The classical algebra  $\W_{N/M}$
has
currents of spins $2,2+M,2+2M,2+3M, \dots ,N$ and
$\W_{\infty/M}$
has
currents of spins $2,2+M,2+2M,2+3M, \dots ,\infty$ [\hee]
and these algebras are not limits of  quantum \W-algebras in general.
There is an algebra \W$G$ associated with any Lie group $G$ and the algebra
associated with $SU(N)$ is the $\W_N$ algebra refered to above [\fatty,\bil].

\chapter{Field Theory Realisations of \W-Algebras}

Consider a theory of
 $D$ free scalar fields $\ffi$ $(i=1,\dots,D)$
with action
$$S_{0}=\ix \partial_{+}\phi^{i}\partial_{-}\phi
^{i}
\eqn\free$$
where the two-dimensional space-time has null coordinates
$x^\mu=(x^+,x^-)$ which are related to the usual Cartesian
coordinates by $x^\pm ={1 \over \sqrt 2}( x^0 \pm x^1)$.
The stress-energy tensor
$$T_{++}={1 \over 2}
\partial_{+}\phi^{i}\partial_{+}\phi^{i}\eqn\tis$$
 is conserved,
$\partial_{-}T_{++}=0$, and  generates the   Poisson bracket algebra
\con\ with $c=0$ (in a canonical treatment regarding $x^-$ as time [\wittt])
which is the conformal algebra with vanishing central charge.
For any   rank-$s$ constant symmetric tensor $d_{i_1 i_2 \dots i_s}$
one can construct a current
$$W_{++\dots +}= {1 \over s}
d_{i_1 i_2 \dots i_s}\partial_{+}\phi^{i_1}\partial_{+}\phi^{i_2}
\dots
\partial
_{+}\phi^{i_s} \eqn\freecur$$
which is conserved, $\partial _-W=0$, and which is a spin-$s$ classical
primary field,
\ie\ its Poisson bracket with $T$ is given by \prim.
The Poisson bracket
of two $W$'s is \walg,
where $\Lambda $ is given by
 $$\Lambda ={1 \over 4  \kappa}
{d_{i \dots j}}^{  m}
d_{k \dots lm} \dpl \ffi  \dots \dpl \fj \dpl \fk \dots \dpl \phi ^l
\eqn\ewrte$$
(the indices $i,j,\dots$ are raised and lowered with the
flat metric $\delta_{ij}$).

Consider first the case $s=3$. In general, closing the algebra generated
by $T,W$ will lead to an infinite sequence of currents $(T,W, \Lambda ,\dots)$.
However, if  $
\Lambda = T^2 $,
for some constant $\kappa$, then the algebra closes non-linearly   on
$T$ and $W$, to give the classical   $\W_3$-algebra
of the last chapter.

In [\he], it was shown that for any number $D$ of bosons, the
necessary and sufficient condition for \lamis\ to be satisfied and hence
for the classical $\W_3$ algebra to be generated
is that the \lq structure constants' $d_{ijk}$ satisfy
$$d^{\ \ \ m}_{(ij}d_{k)lm}=\kappa \delta_{(ij}\delta_{k)l},
\eqn\did$$
This rather striking algebraic constraint has an interesting algebraic
interpretation.\foot{This identity has in fact occurred at least once
before in the physics literature, in the study of five-dimensional
supergravity theories [\gun].} It implies that the $d_{ijk}$
are   the structure constants for a Jordan algebra (of degree $3$)
[\romans] which is a commutative algebra for which  \did\ plays the role of
the Jacobi identities. Moreover, the set of all such algebras has been
classified [\schafer], allowing one to write down the general solution to \did\
[\romans]. In particular, \did\ has a solution for any number $D$ of
bosons. Examples of solutions to \did\ are given for $D=1$ by
$d_{111}=\kappa$ and  for $D=8$ by taking  $d_{ijk}$ proportional to the
$d$-symbol for the group $SU(3)$ [\he]. For $D=2$, the construction of
[\fat] gives a solution of \did\ in which
   the only non-vanishing components
of $d_{ijk} $ are given by $d_{112}=-\kappa $ and $d_{222}=\kappa$, together
with those related to these by symmetry.

 The conserved currents $T,W$ correspond to the invariance of the
free action $S_0$ under the conformal symmetries
$$\delta\phi^{i}=k_{-}\partial_{+}\phi^{i}+\lambda_{--}d^{i}_{\
jk}\partial _{+}\phi^{j}\partial_{+}\phi^{k}
\eqn\sym$$
where the parameters satisfy
$$\partial_{-}k_{-}=0,\qquad \partial_{-}\lambda_{--}=0.
\eqn\concon$$
Symmetries of this kind whose parameters are only functions of $x^+$ (or
only of $x^-$)  will be referred to here as semi-local.
The   symmetry   algebra closes to give
$$\left\lbrack\delta_{k_{1}}+\delta_{\lambda_{1}},\delta_{k_{2}}+\delta
_{\lambda_{2}}\right\rbrack=
\delta_{k_{3}}+\delta_{\lambda_{3}}
\eqn\alg$$
where
$$\eqalign {k_{3}&=\left\lbrack
k_{2}\partial_{+}k_{1}+4 \kappa (\lambda
_{2}\partial_{+}\lambda_{1})T_{++}\right\rbrack\ -\ (1\leftrightarrow2)
\cr
\lambda_{3}&=\left\lbrack2\lambda_{2}\partial_{+}k_{1}+k_{2}\partial
_{+}\lambda_{1}\right\rbrack\ -\ (1\leftrightarrow2)
\cr}\eqn\param$$
In particular, the commutator of two $\lambda$ transformations is a
field-dependent $k$-transformation, which is precisely the transformation
generated by the spin four current $\Lambda=TT$.
The gauge algebra structure \lq constants'
are not constant but depend on the fields $\phi$ through the current $T$,
reflecting the $TT$ term in the current algebra.

The situation is similar for $s>3$. The algebra will close, \ie\ \lop\ will be
satisfied, if the $d$-tensor in \freecur\ satisfies a quadratic constraint
[\hee] and
again this constraint has an algebraic interpretation [\hee].
The $k$ and $\lambda$-transformations become
$$\delta\phi^{i}=k\partial_{+}\phi^{i}+\lambda d^{i}_{\
i_{1}...i_{s-1}}\partial _{+}\phi^{i_{1}}...\partial_{+}\phi^{i_{s-1}}
\eqn\wtr$$
where the parameters  satisfy
 $\partial_{-}k=0$, $ \partial_{-}\lambda=0
$. The symmetry algebra again has field dependent structure \lq constants'.

More generally, any set of constant symmetric tensors
 $d^A_{ij \dots k}$ labelled by some index $A$ can be used to
  construct a set of conserved currents
$$W^A_{++\dots +}=
d^A_{ij \dots k}\partial_{+}\phi^{i}\partial_{+}\phi^j
\dots
\partial
_{+}\phi^k\eqn\freecurn$$
 which are classical
primary fields,
\ie\ their Poisson bracket with $T$ is   \prim.
The current algebra will close if the $d^A$ tensors satisfy certain algebraic
constraints and the Jacobi identities will automatically be satisfied as the
algebra occurs as a symmetry algebra. In this way, a large class of classical
\W-algebras can be constructed by seeking $d^A$-tensors satisfying the
appropriate constraints.
$D$ boson realisations of the $\W_N$ algebras were constructed in this way in
ref. [\hee], where it was shown that the $\W_N$ $d$-tensor constraints had an
interpretation in terms of Jordan algebras of degree $N$, and this again
allowed the explicit construction of solutions to the
$d$-tensor constraints.
These realisations of classical $c=0$ algebras can be generalised to ones
with $c>0$ by introducing a background charge $a_i$, so that the stress tensor
becomes $T=  \dpl \ffi \dpl \ffi + a_i \dpl ^2 \ffi$
 and adding  appropriate higher derivative terms
(\ie\ ones involving $\dpl^m \ffi$ for $m>1$) to the other currents.
The clasical central charge becomes $c=a^2/12$, and for the $N-1$ boson
realisation
of $\W_N$, the structure of the higher derivative terms in the currents
$W_n$ can be derived using Miura transform methods [\bil,\fatty].

Another important realisation of classical \W-algebras is as the Casimir
algebra  of   Wess-Zumino-Witten (WZW) models [\heee]. For the WZW model
corresponding
to a group $G$, the Lie-algebra valued currents $J_+=g^{-1}\dpl g$ generate a
Kac-Moody algebra and are (classical) primary with respect to the Sugawara
stress-tensor $T= {1 \over 2} tr (J_+J_+)$.
Similarly, the higher order Casimirs allow a generalised Sugawara
construction of higher spin currents $tr(J_+^n)$.
For example, for $G=SU(N)$  the set of currents
$W_n= {1 \over n }
tr(J_+^n)$ for $n=2,3,\dots ,N$ generate a closed algebra which is a
classical $\W_N$ algebra [\heee]; similar results hold  for other groups.
Quantum mechanically, however, the Sugawara expressions for the currents  need
normal ordering and must be rescaled [\thierrya,\bais]. For example, in
 the case of $SU(3)$, the quantum
Casimir algebra  leads to a closed \W-algebra (after a certain
truncation) only in the case in which the Kac-Moody algebra is of level
one [\bais].

\W-algebras also arise as symmetry algebras of many other field theories,
including Toda-theories [\bil], free-fermion theories [\heee] and non-linear
sigma-models [\he,\howpapa], giving corresponding realisations of \W-algebras.

\chapter{Gauging \W-Algebras}

In this chapter, the gauging of \W-algebras will be discussed, that is the
coupling of a \W-conformal field theory to \W-gravity gauge fields so that
the extended conformal symmetry is promoted to a local \lq \W-diffeomorphism'
symmetry.
Consider first the simple example of the
free scalar field theory with action  $S_0$ \free\ which is invariant under the
chiral   $\W_3$ transformations \sym\ generated by the currents
$T_{++}= {1\over 2} \dpl \ffi \dpl \ffi $ and $W_{+++}=
{1\over 3} \dpl \ffi \dpl \fj \dpl \fk$, where the tensor $d_{ijk}$ satisfies
\did\ so that the symmetry algebra closes to give \alg,\param.
If the symmetry  parameters $k_-,\lambda_{--}$ are taken to be local, \ie\
the conditions \concon\ are dropped, then the action \free\ varies under \sym\
to give
$$
\delta S_0=\ix (T_{++}\dmi k_-+W_{+++}\dmi \lambda_{--})
\eqn\zar$$
and this can clearly be cancelled by introducing
 gauge fields $h=h_{--},B=B_{---}$ transforming as
$$\delta h_{--}=\dmi k_-+\dots,\qquad
\delta B_{---}= \dmi \lambda_{--} +\dots
\eqn\erter$$
and adding to the action \free\ the Noether coupling [\he]
 $$S_{1}=-\ix \left(h_{--} T_{++} +B_{---} W_{+++} \right ).
\eqn\noeth$$
The action is then invariant to lowest order in the gauge fields.
Remarkably,   terms of higher order in the gauge fields can be added to the
transformations \erter\ in such a way that the linear action $S_0+S_1$
is  fully gauge invariant; surprisingly, no non-linear terms are needed in
the action [\he].
The total action $S=S_0+S_1$ is
invariant under \sym\ together with
 $$\eqalign{
\delta h_{--}=&
\partial_{-}k_{-}+k_{-}\partial_{+}h_{--}-h_{--}\partial
_{+}k_{-}
\cr &
+2 \kappa
\left(\lambda_{--}\partial_{+}B_{---}-B_{---}\partial_{+}\lambda
_{--}\right)T_{++} \cr
\delta
B_{---}= &\partial_{-}\lambda_{--}+2\lambda_{--}\partial_{+}
h_{--}-h_{--}\partial
_{+}\lambda_{--}
\cr &
-2B_{---}\partial_{+}k_{-}+k_{-}\partial_{+}B_{---}.
\cr}
\eqn\vary$$
This is the action for the coupling of the free scalar field theory to chiral
$\W_3$ gravity [\he]. The gauge fields are Lagrange multipliers imposing the
constraints $T=0$, $W=0$ and these constraints satisfy the algebra
\con,\prim, \walg, \lamis.
The gauge algebra
is still given by \alg\ and \param, up to terms that vanish when the
classical equations of motion are satisfied.
However, as $T=0$ is   the $h_{--}$ equation of motion, the symmetry algebra
\alg\ is equivalent on-shell to the simpler one obtained by setting $T=0$ in
\alg, so that two $\lambda$ transformations commute up to equations of motion.
This simpler algebra corresponds to the commutation relations \con,\prim\ and
$$
\left\lbrack
W(x^+),W(y^+)\right\rbrack  =0
  \eqn\zamone$$
While the previous algebra given by \con,\prim, \walg, \lamis\ only had a
maximal finite subalgebra of $SL(2,\IR)$ (generated by the modes $L_0,
L_{\pm 1}$), the algebra in which \walg\ is replaced by \zamone\ has an
interesting finite subgroup of $ISL(2,\IR)$, a group contraction of
$SL(3,\IR)$ (generated by the modes
$L_0, L_{\pm 1},W_0,W_{\pm 1},W_{\pm 2}$).

So far only a right-handed \W -algebra has been gauged.
Gauging the left-handed reparameterisations generated by $T_{--}=\half \dmi
\ffi \dmi \ffi$ by introducing a gauge field $h_{++}$ leads to a \lq
heterotic' model in which a right-moving $\W _3$ algebra and a left-handed
Virasoro algebra are gauged [\he].
To gauge both a left-handed and a right-handed $\W _3$ algebra requires a
further spin-three gauge field $B_{+++}$ corresponding to the current
$W_{---}= {1 \over 3 }
d_{ijk}\dmi \ffi \dmi \fj \dmi \fk$. The action is given by
$S=S_0 +S_1 +\dots$ plus terms of quadratic order and higher in the gauge
fields, where $S_1$ is the Noether coupling
$$S_{1}=-\ix\left( h_{--}T_{++}+  h_{++}T_{--}+
 B_{---}W_{+++} +
 B_{+++}W_{---}\right )
\eqn\noether$$
The full action can then be constructed iteratively using the Noether
method. However, this requires an infinite number of steps as the action is
non-polynomial in   $h_{\pm \pm}$ and in $B_{\pm \pm \pm } \partial _{\mp}
\ffi$, both of which have zero (world-sheet) dimension (as $B_{\pm \pm \pm
}$ has dimension $-1$ and $\partial \phi$ has dimension $+1$) [\he].

The gauge fields $h^{\pm \pm},B^{\pm \pm \pm}$ are  the
components of  traceless
  gauge fields
$ h^{\mu \nu}, B^{\mu \nu \rho}$ and the action given by adding \noether\
to $S_0$ can be rewritten as
 \snoth. The tracelessness condition on the gauge fields can be dropped and
traces $h^{+-},B^{\pm +-}$ formally introduced. However, they were not
needed for the linearised action, which is consequently invariant under the
linearised \W-Weyl transformations \wweyl, or equivalently
$\delta h^{+-}=\Omega ,\delta B^{\pm +-}=\Omega^\pm$. Indeed, the traces are
not needed at all in the classical theory, \ie\ there is a non-linear form of
the \W-Weyl symmetry. However, if there are anomalies, this need not be the
case in the quantum theory.

This example is typical of the general case.
Consider any classical conformal field theory with a \W-algebra symmetry,
(e.g. a free boson or fermion model, a non-linear sigma model, a
Wess-Zumino-Witten model, a coset model, a Toda model,...). Let the classical
action be $S_0$ and the  currents generating the \W-algebra be
$W_{+A}, W_{-A}$ labelled by some index $A$ and  satisfying the chiral
conservation laws
$\dmi W_{+A}=0$, $\dpl W_{-A}=0$. To gauge a chiral \W-algebra, say the
right-handed one generated by the currents $W_{+A}$ alone, a gauge field
$h^{+A}$ is introduced for each current and  adding the  Noether coupling
gives the linearised action
$$S=S_0+\sum_A \ix h^{+A} W_{+A}
\eqn\newt$$
It was seen in the case of chiral $\W_3$ that no higher order terms are
needed and that this action was fully gauge invariant. Remarkably, it can be
shown that this is the case in general and that for any classical conformal
field theory and any chiral \W-algebra, the full gauge-invariant action is
linear in the gauge fields and given simply by the Noether coupling [\heee].
To gauge both left and right-handed \W-algebras, one introduces gauge fields
$h^{+A}$ and $h^{-A}$. In this case, adding the Noether coupling does not
give a gauge-invariant theory in general and the full theory is
non-polynomial in the gauge fields. To lowest order it is given by
$$S=S_0+\sum_A \ix \bigl[
h^{+A} W_{+A}+ h^{-A} W_{-A}+O(h^2)\bigr]
\eqn\newtq$$
and the higher order corrections can be calculated to any given order in the
gauge fields using the Noether method. Nevertheless, it will be seen in the
next chapter, that the non-chiral gaugings can be written in a type of
canonical formalism in which the gauged action is given by a Noether coupling.
However, the  canonical momenta cannot be eliminated in closed form
so that this approach only gives an implicit form of the action.

\chapter{Canonical Construction of Non-Chiral \W-Gravity}

We now turn to the full non-linear structure of the non-chiral \W-gravity
coupling, and again the example of the free boson realisation of $\W_3$ will
be used. There are two canonical-style approaches which give a simple but
implicit form of the action. The first is a conventional Hamiltonian treatment
[\mikoham]. The free action \free\ can be written in first-order form as
$$S_{0}=\ix  \left\lbrack\pi_{i}\partial_{\tau}\phi^{i}-{1\over
2}\pi_{i}\pi^{i}-{1\over2}\mathop{\partial_{\sigma}\phi^{i}\partial
}\nolimits_{\sigma}\phi^{i}\right\rbrack
\eqn\firor$$
with $ \tau=x^0, \sigma=x^1$.
The momentum $\pi^i$ is an auxiliary field that can be eliminated using its
equation of motion $\pi^i=\partial _\tau \ffi$ to recover \free.
The currents defined by
$$T_{\pm\pm}\hbox{(}\Pi\hbox{)=}{\hbox{1}\over\hbox{
2}}\Pi^{i}_{\pm}\Pi^{i}_{\pm},
\qquad W_{\pm\pm\pm}\hbox{(}\Pi\hbox{)=}{\hbox{1}\over
\hbox{3}}d_{ijk}\Pi^{i}_{\pm}\Pi^{j}_{\pm}\Pi^{k}_{\pm}
\eqn\curpi$$
where
$$\Pi^{i}_{\pm}\hbox{=}{\hbox{1}\over\sqrt{\hbox{
2}}}\left(\pi^{i}\pm\partial_{\sigma}\phi^{i}\right)\eqn\piis$$
generate a Poisson bracket algebra consisting of two copies of the classical
$\W_3$ algebra given in chapter 2. The first-class constraints
$$T_{\pm\pm}\hbox{(}\Pi\hbox{)}\sim\hbox{0
},\qquad W_{\pm\pm\pm}\hbox{(}\Pi\hbox{)}\sim\hbox{0}
 \eqn\coco$$
can then be imposed using Lagrange multipliers $h^{\pm\pm},B^{\pm\pm\pm}$,
so that the action is given by
$$ S{=}S_{ {0}} {-}\ix \left
\lbrack h^{ {++}}T_{ {++}} {(}\Pi
 {)+}h^{ {--}}T_{ {--}} {
(}\Pi {)+}B^{ {+++}}W_{ {+++}} {
(}\Pi {)+}B^{ {---}}W_{ {---}} {
(}\Pi {)}\right\rbrack
\eqn\hamact$$
Here, $S_0$ is the free action given by \firor.
After shifting
  the fields $h^{\pm\pm} \rightarrow h^{\pm\pm}+1$,
 $$ \eqalign{
 S=&\ix
\bigl[
\pi_{i}\partial_{\tau}\phi^{i}
-h^{ {++}}T_{ {++}} {(}\Pi
 {)-}h^{ {--}}T_{ {--}} {
(}\Pi  )
\cr
 &- B^{ {+++}}W_{ {+++}} {
(}\Pi {)-}B^{ {---}}W_{ {---}} {
(}\Pi {)}
\bigr]\cr}
\eqn\soso$$
This gives the complete coupling of scalar fields to non-chiral $\W_3$
gravity in Hamiltonian form. However, although the field equations for the
momenta $\pi^i$ are still algebraic,
they are difficult to solve  in closed form, so that eliminating the momenta is
problematic.
 However, they can be solved to  any given order in the
gauge fields, and the solution can then be used to find the Lagrangian form
of the action to that order in the gauge fields.

The action given by \hamact,\soso\ is an example of a first class action of
the form $$S {=}\intt d\tau\ \left\lbrack
p_{a}\partial_{\tau}q^{a} { -}\lambda^{A}G_{A}\right\rbrack
\eqn\hamo$$
where the coordinates $q^a(\tau)$ and momenta $p_a(\tau)$
correspond to $\ffi (\sigma,\tau)$ and $\pi_i(\sigma,\tau)$ respectively,
with the index $a$ representing both the discrete index $i$ and the
continuous variable $\sigma$, so that summation over $a$
corresponds to summation over $i$ and integration over $\sigma$.
The Lagrange multipliers $\lambda^{A}(\tau)$ correspond to
$h^{\pm\pm}(\sigma),B^{\pm\pm\pm}(\sigma)$ and impose the constraints $G_A
\sim 0$, corrsponding to \coco.
Suppose the  Poisson bracket  algebra generated by the constraints closes to
give
 $$\left\lbrace
G_{A} {,}G_{B}\right\rbrace {=}{f_{AB}}^{  C}G_{C}
\eqn\curalg$$
 for some ${f_{AB}}^{  C}$, which may depend on the phase space
variables. In our example, \curalg\ is just the classical $\W_3$ algebra.

Any action of the form \hamo\ where the constraints satisfy \curalg\ is
invariant
under the following local symmetries with parameter $\alpha ^A(\tau)$
$$\delta p_{a} {=}\alpha^{A}\left\lbrace
G_{A} { ,}p_{a}\right\rbrace, \qquad \delta
q^{a} {=}\alpha^{A}\left\lbrace G_{A} { ,}q^{a}\right\rbrace
\eqn\kjef$$
$$\delta\lambda^{A} {=}\partial_{\tau}\alpha^{A} {
-}{f_{BC}}^{  A}\lambda^{B}\alpha^{C}\eqn\varoojdf$$
These then give  the $\W_3$ gravity symmetries of the  action \soso, which are
given explicitly in [\mikoham].

The Hamiltonian approach has the disadvantage that   two-dimensional
Lorentz covariance is not manifest. A related method which does maintain
covariance was found  by Schoutens, Sevrin and van Nieuwenhuizen in
[\van].  This method, which was  found before the Hamiltonian one
described above, was motivated by a careful study of the terms that occur in
the order-by-order construction of the action [\van].
Instead of a single momentum $\pi^i$ conjugate to each $\ffi$, a covariant
vector $\pi ^i_{\mu}$ is introduced.
The free action $S_0$ can then be written as
$$
S=\ix\Bigl(-{1\over2}\partial_{+}\phi^{i}\partial_{-}\phi
^{i}-\pi ^{i}_{+}\pi ^{i}_{-}+\pi ^{i}_{+}\partial_{-}\phi^{i}+\pi
^{i}_{-}\partial
_{+}\phi^{i} \Bigr)
\eqn\vree$$
The $\pi$ field equation is algebraic:
$$
{\delta S_0 \over \delta \pi ^{i}_{\pm}}=0 ~ \Rightarrow ~
\pi ^{i}_{\pm}=\partial _{\pm} \ffi
\eqn\erfbfg$$
so that the $\pi ^i_{\mu}$ are auxiliary fields.
Using \erfbfg\ to eliminate the auxiliary fields (\ie\
setting $ \pi ^{i}_{\pm}=\partial _{\pm} \ffi$), the action reduces to the
usual free form \free.  The next step is to introduce \lq $\pi$-currents'
$T(\pi),W(\pi)$
$$
T_{\pm \pm}(\pi)= {1 \over 2} \pi^i_\pm \pi^i_\pm
\qquad
W_{\pm \pm \pm}(\pi)= {1 \over 3}d_{ijk} \pi^i_\pm \pi^j_\pm \pi^k_\pm
\eqn\pcurr$$
Adding the Noether coupling
$$S_{1}=-\ix\Bigl[ h_{--}T_{++}(\pi)+  h_{++}T_{--}(\pi)+
 B_{---}W_{+++}(\pi) +
 B_{+++}W_{---}(\pi)\Bigr]
\eqn\noetherpi$$
to \vree\ imposes the constraints that the $\pi$-currents vanish.

This Noether coupling turns out to be all that is needed to give a
gauge-invariant theory. The complete action is then [\van]
 $$
\eqalign {
S=&\ix\Bigl(-{1\over2}\partial_{+}\phi^{i}\partial_{-}\phi
^{i}-\pi ^{i}_{+}\pi ^{i}_{-}+\pi ^{i}_{+}\partial_{-}\phi^{i}+\pi
^{i}_{-}\partial
_{+}\phi^{i}
\cr &
-{1\over2}h_{--}\pi ^{i}_{+}\pi ^{i}_{+}
-{1\over2}h_{++}\pi ^{i}_{-}\pi ^{i}_{-}
\cr &
-{1\over
3}B_{+++}d_{ijk}\pi ^{i}_{-}\pi ^{j}_{-}\pi ^{k}_{-}
-{1\over3}B_{---}d_{ijk}\pi ^{i}_{+}\pi ^{j}_{+}\pi ^{k}_+
\Bigr)
\cr }
\eqn\one
$$
(This form of the action is a generalisation of one given in [\pop]
and is related to the    action of [\van] by field redefinitions.)
This action is invariant under the diffeomorphisms and
$\lambda$-transformations
$$\eqalign {
\delta\phi^{i}=&k_{-}\pi ^{i}_{+}+\lambda_{--}d^{i}_{\
jk}\pi ^{j}_{+}\pi ^{k}_{+}+(+\leftrightarrow -)
\cr \delta h_{\pm\pm}=&
\partial_{\pm}k_{\pm\ }+k_{\pm }\partial
_{\mp}h_{\pm\pm}-h_{\pm\pm}\partial  _{\mp}k_{\pm }
\cr &
+2\kappa T_{\mp\mp}( \Pi)\left
(\lambda_{\pm\pm}\partial  _{\mp}B_{\pm\pm\pm}-B_{\pm\pm\pm}\mathop{\partial
 _{\mp}\lambda}\nolimits_{\pm\pm}\right)
\cr
\delta B_{\pm\pm\pm}=&
\partial_{\pm}\lambda_{\pm\pm}+2\lambda_{\pm
\pm}\partial  _{\mp}h_{\pm\pm}-h_{\pm\pm}\partial  _{\mp}\lambda
_{\pm\pm}
\cr &
-2B_{\pm\pm\pm}\partial  _{\mp}k_{\pm }+k_{\pm}\mathop{\partial
  _{\mp}B}\nolimits_{\pm\pm\pm}
\cr
\delta \pi ^{i}_{\pm}=&\ \partial_{\pm}\left(k_{\mp}\pi ^{i}_{\pm}+\lambda
_{\mp\mp}d^{i}_{\ jk}\pi ^{j}_{\pm}\pi ^{k}_{\pm}\right)
\cr}
\eqn\clevvar$$
where $T_{\pm \pm}( \Pi) = \half \pi ^{i}_{\pm} \pi ^{i}_{\pm}$.
 The field equation for the auxiliary fields is algebraic
$$\pi ^{i}_{\pm}=\partial_{\pm}\phi^{i}-h_{\pm\pm}\pi ^{i}_{\mp}-B_{\pm\pm
\pm}d^{i}_{\ jk}\pi ^{j}_{\mp}\pi ^{k}_{\mp}
\eqn\aux$$
but difficult to solve for $\pi$ in closed form, so that again there is not a
closed form for the action without $\pi$'s. Nevertheless, one can solve for
$\pi$ to any given order in the gauge fields and the result agrees with that
obtained by using the Noether method to find the corrections
to \free,\noether\
to that order in the gauge fields.

To obtain a better understanding of these actions, it may be useful to
consider  setting the spin-three gauge-fields
to zero in the actions considered above to obtain the coupling to pure
(spin-two) gravity, which can then be compared with the conventional
minimal coupling to gravity.
The Noether coupling approach gives the action
$$
S_{n}=\ix \left( \dpl \ffi \dmi \ffi -{1\over 2}h_{++}  \dmi \ffi \dmi \ffi
-{1\over 2} h_{--}\dpl \ffi \dpl \ffi+ O(h^2) \right)
\eqn\nonon$$
Although one could calculate some of the higher order corrections to this
and attempt to guess the general form, this is clearly not the best way of
finding the coupling of a scalar field to gravity.
The approach of [\van] gives
 $$
\eqalign {
S=&\ix\Bigl(-{1\over2}\partial_{+}\phi^{i}\partial_{-}\phi
^{i}-\pi ^{i}_{+}\pi ^{i}_{-}+\pi ^{i}_{+}\partial_{-}\phi^{i}+\pi
^{i}_{-}\partial
_{+}\phi^{i}
\cr &
-{1\over2}h_{--}\pi ^{i}_{+}\pi ^{i}_{+}
-{1\over2}h_{++}\pi ^{i}_{-}\pi ^{i}_{-}
\Bigr)
\cr }
\eqn\pone
$$
and the $\pi$ field equation  is
$$\pi ^{i}_{\pm}=\partial_{\pm}\phi^{i}-h_{\pm\pm}\pi ^{i}_{\mp}
\eqn\aux$$
which can   be solved explicitly to give
$$
\pi ^{i}_{\pm}={\partial _\pm \ffi -h_{\pm \pm}\partial _\mp \ffi
\over 1-h_{--}h_{++}}
\eqn\sole$$
Substituting \sole\ into \pone\ gives
the complete non-polynomial form of the action
$$
S=\ix {1 \over 1-h_{++}h_{--}}\Bigl[
(1+h_{++}h_{--}) \dpl \ffi \dmi \ffi - h_{--}T_{++}-h_{++}T_{--}
\Bigr]
\eqn\kshg$$
This gives the full non-linear corrections to \nonon.

Similarly, the Hamiltonian approach gives
$$S =\ix  \left\lbrack\pi_{i}\partial_{\tau}\phi^{i}
-h^{ {++}}T_{ {++}} {(}\Pi
 {)-}h^{ {--}}T_{ {--}} {
(}\Pi {)}
\right\rbrack
\eqn\firorgr$$
and the momenta can again be eliminated explicitly to
give a non-polynomial action similar to \kshg.

Of course, most people
would prefer to use a little geometry and write down the standard minimal
coupling to a metric $g_{\mu \nu}$
$$S= {1 \over 2} \ix \sqrt{-g} g^{\mu \nu}
\dm \ffi \dn \ffi
\eqn\mincoup$$
If one chooses to parameterise the metric as
$$
g_{\mu\nu}=\Omega\left(\matrix{2h_{++}&1+h_{++}h_{--}\cr\noalign{\smallskip}
1+h_{++}h_{--}&
2h_{--}\cr
}\right)
\eqn\two
$$
then $\Omega$ drops out
of the action \mincoup\ as a consequence
of classical Weyl invariance and \mincoup\ becomes
precisely \kshg. Note that, contrary to claims sometimes made, \two\ does not
correspond to partial gauge-fixing; \two\ is simply a convenient
parameterisation of a {\it general} metric $g_{\mu \nu}$.

However, for non-zero $B$, \aux\ gives an   equation for $\pi$ which is
difficult to solve in closed form,
 although it is straightforward to solve
order by order in $B$; substituting the perturbative solution to \aux\ back
into the action
recovers the   results of the Noether method. However, just as the
non-linearity in $g_{\mu \nu} $ is best understood in terms of Riemannian
geometry, it seems likely that the non-linearity in $B$ can also be best
understood in terms of some higher spin geometry, which would also allow
the coupling of \W -gravity  to more general matter systems.

The two canonical approaches described here
for $\W_3$-gravity
also work for other \W-algebras and other matter systems. The Hamiltonian
approach clearly works quite generally; one writes the matter action in first
order form and replaces time derivatives of fields in the \W-currents by the
corresponding momenta. The full action is given by adding the Noether
coupling of these currents to Lagrange multiplier gauge fields. The resulting
action is of the form \hamo\ and so invariant under the transformations \kjef,
\varoojdf.
The covariant canonical approach of [\van] has been generalised to
$w_\infty$ [\pop], $W_N$ [\hee] and indeed free boson
realisations of {\it any} \W-algebra [\hee]. It has also been applied to
non-linear sigma-models, free-fermion models [\hee]
 and supersymmetric models [\basti,\miko] and
probably again  applies quite generally.

\chapter{The Geometry of \W-Gravity}

The non-polynomial structure of gravity is best understood in terms of
Riemannian geometry and this suggests that the key to the  non-linear
structure of \W-gravity might   be found in some higher spin generalisation
of Riemannian geometry. The approaches of [\van] and [\mikoham] describe
\W-gravity in an
implicit form, but an understanding of the  non-linear
structure of the theory without auxiliary fields seems desirable. A
\lq covariantisation' of the approach of [\van] is given in [\vann,\popp],
 but the
resulting theories still have auxiliary fields. Other approaches to the
geometry of \W-gravity are presented in [\wit,\wot,\sot]. In [\hegeom], a
geometric formulation
of $w_\infty$ gravity was derived and this will now be briefly reviewed.

 Riemannian geometry  is based on a line element
$ds=( g_{ij} d \phi ^i d  \phi ^j
)^{1/2}$, while a  spin-$n$ gauge field could be
used to define a geometry based on a line element
$
ds=(g_{i_1 i_2 \dots i  _n} d \phi^{i_1} d \phi^{i_2}\dots d
\phi^{i_n})^{1/n} $ (first considered by Riemann [\riemy]).
A further generalisation is to consider a line element
$
ds=N(\phi,d\phi)
 $
where $N$ is some function which is required to satisfy the homogeneity
condition
$
N(\phi,\lambda d\phi)= \lambda N(\phi,d\phi)
 $. This defines a Finsler geometry [\finre] and generalises the
spin-$n$ Riemannian line element.
To describe \W-gravity, it seems appropriate to generalise still further
and drop the homogeneity condition on $N$, so that the line element can be
written as
$$
ds^2={\cal N}[\ffi, d \ffi]= g_{ij}(\phi) d \ffi d \fj+
d_{ijk}(\phi) d \ffi d \fj d \fk +\dots
\eqn\fififi$$
One can then contemplate  the extension of the diffeomorphism symmetry
$\delta \ffi=\Lambda ^i(\phi)$ to the action of a much larger group,
consisting of transformations of the form $\delta \ffi=\Lambda
^i(\phi,d\phi)$. Given a  scalar field  $\ffi(x^\mu)$ taking values in this
space, the line element  can be pulled back to the \lq world-sheet' to give
the world-sheet line-element
$$\eqalign{ds^2=&
{\cal N}^*[\ffi(x), \dm \ffi dx^\mu] = g_{ij}(\phi(x)) \, \dm \ffi \dn \fj
\, dx^\mu dx^\nu
\cr &+
d_{ijk}(\phi(x))  \, \dm \ffi \dn \fj \dr \fk \, dx^\mu dx^\nu dx^\rho +\dots
\cr }
\eqn\fififo$$
and the transformations of $\phi$ now take the form
$$\delta \ffi=\Lambda
^i(\phi,\dm\phi)\eqn\wutwf$$
In fact, these transformations are too general and as will be seen, it is
necessary to restrict to a subgroup of these. To  use the line element
\fififo\ to construct an action, it is now necessary to introduce
world-sheet gauge fields $g^{\mu \nu } ,B^{\mu \nu \rho},\dots$.
This is best done by introducing a generating function
$$
F(x^\mu , y_\mu)=
g^{\mu \nu } (x) y_\mu y_\nu +B^{\mu \nu \rho}(x) y_\mu y_\nu y_ \rho +\dots
\eqn\gag$$
 where $y_\mu$ is some vector on the world-sheet $M$.
Note that \gag\ is a generalisation of the inverse metric on $M$ in the
same sense that \fififi\ is a
generalisation of the   metric on the target space.
Note that $x^\mu,
y_\mu$ can be thought of as coordinates on the cotangent bundle of the
world-sheet, $T^* M $, and if $F$ is required to be a function on $T^*M$,
then the gauge fields $g^{\mu \nu } ,B^{\mu \nu \rho},\dots$ are
tensor fields on $M$.
The function $F$ can be used to define some $\tilde F$ given by
$$
\tilde F(x^\mu , y_\mu)=
\tilde g_{(2)}^{\mu \nu } (x) y_\mu y_\nu +\tilde g_{(3)}^{\mu \nu \rho}(x)
y_\mu y_\nu y_ \rho +\dots \eqn\gagti$$
where
$$\tilde g_{(2)}^{\mu \nu } =\sqrt {-g}g^{\mu \nu }, \qquad
\tilde g_{(3)}^{\mu \nu
\rho} =\sqrt {-g}(B^{\mu \nu
\rho}-{3 \over 2}g^{(\mu \nu}B^{\rho) \sigma \tau}g_{\sigma \tau}), \dots
\eqn\tist$$
are tensor densities on $M$.
Then the \W-gravity action is given by the natural product
$$\eqalign{S&=\ix \tilde F \cdot \cal{N}^*
\cr &= \ix \left[
g_{ij}(\phi(x)) \dm \ffi \dn \fj
\tilde g^{\mu \nu }_{(2)}
+
d_{ijk}(\phi(x))  \dm \ffi \dn \fj \dr \fk \tilde g_{(3)}^{\mu \nu
\rho}+\dots \right]
\cr}
\eqn\hjsv$$

To see how these structures arise from a slightly different viewpoint,
consider the simple example of the realisation of the $w_\infty$ in terms
of   a free single boson $\phi$ [\pop]. No $d$-tensors are needed, and the
infinite set of currents
$$W_{+n}= {1 \over n} (\dpl \phi)^n, \qquad n=2,3,\dots
\eqn\wiss$$
generate the algebra $w_\infty$ [\winf] (a certain $c=0$ limit of the
$\W_\infty$ algebra of [\qwinf]).
These currents generate the infinitesimal transformations
$$
\delta \phi = \sum _{n=2} ^ \infty
\lambda ^{+n}(x^+)(\dpl \phi)^{n-1} \equiv \Lambda (x^+, \dpl \phi)
\eqn\dph$$
Here $W_{+2}$ is  the stress-tensor and $\lambda ^{+2}(x^+)$
is the parameter of conformal transformations.
The symmetry algebra is
$$\lbrack\delta_{\Lambda},\delta_{\Lambda^\prime}\rbrack=\delta_{\lbrace
\Lambda,\Lambda^\prime\rbrace}\eqn\alg$$
where
$$\lbrace\Lambda,\Lambda^\prime\rbrace={\partial\Lambda\over\partial
x^{+}}{\partial\Lambda^\prime\over\partial y_{+}}-{\partial\Lambda
^\prime\over\partial x^{+}}{\partial\Lambda\over\partial y_{+
}}
\eqn\poiss$$
is the Poisson bracket on the phase space with coordinates $x^+,y_+\equiv
\dpl \phi$. If the world-sheet is cylindrical, then $x^+ $ is the
coordinate for a circle and  $x^+,y_+$ are coordinates for the cotangent
bundle $T^*S^1 \sim S^1 \times \IR$. The symmetry algebra is then isomorphic
to the Poisson bracket algebra on  $T^*S^1$, which is also the algebra of
symplectic diffeomorphisms of $T^*S^1$, \ie\ those diffeomorphisms of the
two-dimensional space $T^*S^1$ that preserve the symplectic structure
$dx^+_\Lambda dy_+$.

The currents $W_{-n}= {1 \over n} (\dmi \phi)^n$ generate a second
commuting  copy of $w_\infty$ so that the symmetry algebra is given by two
copies of the symplectic diffeomorphisms,
$Diff_0(T^*S^1)\times Diff_0(T^*S^1)$. The symmetry algebra
is a subalgebra of the more general set of transformations
$$\delta\phi=\sum^{\infty}_{n=2}
\lambda^{\mu_{1}\mu_{2}...\mu_{n-1}}_{(n)}(x^{\nu
})\partial_{\mu_{1}}\phi\partial_{\mu_{2}}\phi....\partial_{\mu_{n-1}}\phi
\equiv\Lambda(x^{\mu},y_{\mu})
\eqn\sdtrans$$
where
$y_{\mu}=\partial_{\mu}\phi
$
and the $\lambda^{\mu_{1}\mu_{2}...\mu_{n-1}}_{(n)}(x^{\nu
})$ ($n=2,3,\dots$) are infinitesimal parameters which are symmetric
tensor fields on $M$. These transformations satisfy the algebra \alg, where
the Poisson brackets are those for the phase space with coordinates $x^\mu
, y_\mu$, which is the cotangent bundle of the world-sheet, $T^*M$. Thus
\sdtrans\ is a field-theoretic realisation of the symplectic diffeomorphism
algebra, $Diff_0(T^*M)$.
It has been suggested [\wit] that this might be the symmetry algebra for the
corresponding \W-gravity theory, but as will be seen, it is a sub-algebra of
this that is in fact needed.

Before proceding to the full theory, it will be useful to consider first
  linearised $w_\infty$ gravity. The currents $W_{\pm n}$ are the only
non-vanishing components of the  symmetric tensor
 current given by
$$
W_{\mu_1 \mu_2 \dots \mu_n}^n= {1 \over n} \partial _{\mu_1}\phi
\partial _{\mu_2}\phi \dots \partial _{\mu_n}\phi - \ traces
\eqn\wcurr$$
which is   conserved, $\partial ^{\mu_1}
W_{\mu_1 \mu_2 \dots \mu_n}^n=0$ and traceless,
$\eta^{\mu_1 \mu_2}W_{\mu_1 \mu_2 \dots \mu_n}^n=0$. From chapter 1, the
linearised \W-gravity action is given by the Noether coupling.
The action \noeth\ can be rewritten as
$$S=\int d^{2}x\left\lbrack {1\over2}\eta ^{\mu\nu
}\partial_{\mu}\phi\partial_{\nu}\phi+ \sum^{\infty}_{n=2}{1\over
n}\tilde h^{\mu_{1} ...\mu_{n}}_{(n)}\partial_{\mu_{1}}\phi
...\partial_{\mu_{n}}\phi+O(\tilde h^{2})\right\rbrack
\eqn\renoeth$$
where the $\tilde h^{\mu_{1}\mu_{2}...\mu_{n}}_{(n)}$ are symmetric tensor
gauge fields satisfying
 $$\eta_{\mu\nu}\tilde h^{\mu\nu\rho...\sigma}_{(n)}=0+O(h^{2})
\eqn\traa$$
at least to lowest order in the gauge fields.
The transformation of the scalar fields is of the form \sdtrans\
with the parameters satisfying
the tracelessness condition
$$\eta_{\mu\nu}\lambda^{\mu\nu\rho....\sigma}_{(n)}=0 +O(h^{2})
\eqn\trlam$$
at least to lowest order in the gauge fields.

Let us turn now to the full non-linear theory. The action depends  on
 $\dm \phi$ but not on higher order derivatives and so can be written as
 $$S=\int _M d^{2}x\tilde{F}(x,\partial\phi)
\eqn\sis$$
for some $\tilde{F}$, which has the following expansion in $y_\mu =\dm \phi$:
$$\tilde{F}(x,y)=\sum^{\infty}_{n=2}{1\over n} {\tilde{g}}
^{\mu_{1}\mu_{2}...\mu_{n}}_{(n)}\mathop{(x)y}\nolimits_{\mu_{1}}y_{\mu
_{2}}...y_{\mu_{n}}
\eqn\fis$$
where ${\tilde{g}}
^{\mu_{1}\mu_{2}...\mu_{n}}_{(n)}(x)$ are gauge fields
which are tensor densities on $M$. The ${\tilde{g}}
^{\mu_{1} ...\mu_{n}}_{(n)} $ are non-polynomial functions of the
$\tilde h^{\mu_{1} ...\mu_{n}}_{(n)}$ (e.g. $\tilde g^{\mu \nu }_{(2)}=\eta
^{\mu \nu }+\tilde h^{\mu \nu }_{(2)} +O(\tilde h^2)$) but it will be
convenient to work with the $\tilde g$ gauge fields, in terms of which the
action is linear, rather than the $\tilde h$ gauge fields.
The transformation of $\phi$ takes the form \sdtrans.
 From the linearised analysis,
it is seen that $\Lambda$ and $\tilde F$ must be restricted to satisfy
equations which take the following form to lowest order in the gauge fields
$$\eta_{\mu\nu}{\partial^{2}\Lambda\over\partial y_{\mu}\partial y_{\nu
}}=0+\dots , \qquad \eta_{\mu\nu}{\partial^{2}\tilde F
\over\partial y_{\mu}\partial y_{\nu
}}=-2+\dots
\eqn\linlam$$
These impose the tracelessness of the parameters \trlam\  and of the
linearised gauge fields, to lowest order in the gauge fields. The full
theory should have non-linear constraints that generalise these and which
do not involve any background metric $\eta^{\mu \nu}$. This is indeed the
case, and the full constraints take a strikingly simple form:
$$
det \left ({\partial ^2   \tilde F (x,y)\over \partial y_\mu
\partial y_ \nu} \right)=-1
\eqn\decon$$
and
$$
det \left ({\partial ^2   \over \partial y_\mu
\partial y_ \nu} [\tilde F +\Lambda](x,y) \right)=-1
\eqn\lamcon$$
 Expanding   \decon\ in $y_\mu$
gives an infinite number of algebraic constraints on the density gauge fields
$\tilde g_{(n)}^{\mu \nu \dots}$
$$ det \left(\tilde g_{(2)}^{\mu \nu}(x) \right) = -1, \qquad
\tilde g_{(3)}^{\mu \nu \rho } \tilde g_{(2) \mu \nu}=0, \dots
\eqn\geco$$
and these can be solved in terms of unconstrained gauge fields $g^{\mu
\nu}, B^{\mu \nu \rho}, \dots$ to give \tist.
The theory written in terms of these unconstrained gauge fields
is invariant under an infinite set of local symmetries generalising
Weyl symmetry:
$$  \delta
g^{\mu\nu} =\sigma_{(2)}g^{\mu\nu} , \qquad \delta
B^{\mu\nu\rho}_{(3)}=\sigma_{(2)}B^{\mu\nu\rho}_{(3)}+{3 \over 2}\sigma
^{(\mu}_{(3)}g^{\nu\rho)} ,
\dots
\eqn\exweyl$$
The transformations can be written in terms of the generating function \gag\ as
$\delta F(x,y)=\sigma(x,y)F(x,y)$ where
 $\sigma(x,y)=\sigma_{(2)}(x)+\sigma^{\mu}_{(3)}(x)y_{\mu}+...
$
These transformations can be used to remove all traces from the gauge
fields, leaving only   traceless gauge fields, as in [\pop,\vann].

Similarly, expanding   \lamcon\ in $y_\mu$
gives an infinite number of algebraic constraints on the parameters of the
form
$\tilde g_{(2) \mu \nu}\lambda _{(n)}^{\mu \nu \dots}=0+\dots$ and these
can agan be solved in terms of unconstrained quantities [\hegeom].

To summarise, the full non-linear action for $w_\infty$ gravity coupled to
a single boson is given by \sis,\fis\ where
the gauge fields ${\tilde{g}}
^{\mu_{1}\mu_{2}...\mu_{n}}_{(n)}(x)$
are required to satisfy the algebraic constraints given by
expanding \decon. These constraints can be solved in terms of unconstrained
gauge fields as described above, but it is not necessary to do so. The
action is invariant (up to
a surface term) under   $w_\infty$ gravity transformations under which the
scalar fields transform as \sdtrans, where the infinitesimal parameters
are required to satisfy the constraint \lamcon, and this constraint implies
that the symmetry algebra is a subalgebra of $Diff_0(T^*M)$. The gauge
fields  transform as
 $$\eqalign{
\delta\mathop{\tilde{g}}\nolimits^{\mu_{1}\mu_{2}...\mu_{p}}_{(p)}&=\sum
_{m,n=2}^\infty
\delta_{m+n,p+2}\biggl[(m-1)\lambda^{(\mu_{1}\mu_{2}...}_{(m)}\partial
_{\nu}\mathop{\tilde{g}}\nolimits^{...\mu_{p})\nu}_{(n)}-(n-1)\mathop{\tilde
{g}}\nolimits^{\nu(\mu_{1}\mu_{2}...}_{(n)}\partial_{\nu}\lambda^{...\mu
_{p})}_{(m)} \cr &
+{(m-1)(n-1)\over p-1}\partial_{\nu}\left\lbrace\lambda^{\nu(\mu
_{1}\mu_{2}...}_{(m)}\mathop{\tilde{g}}\nolimits^{...\mu_{p})}_{(n)}-\mathop{\tilde
{g}}\nolimits^{\nu(\mu_{1}\mu_{2}...}_{(n)}\lambda^{...\mu_{p})}_{(m)}\right
\rbrace\biggr]
\cr}
\eqn\denvar
$$
{}From \denvar, the $\tilde g_{(s)}$ transform as tensor
densities under reparameterisations of $M$  (\ie\ $\lambda _{(2)}$
transformations), as expected.

The equation  \decon\ is of a type that plays an important role in
geometry.
 Let $\zeta_\mu,
\bar \zeta _{\bar\mu}$ ($\mu =1,2$) be complex coordinates on $\IR ^4$.
Then, for each $x^\mu$, a solution $\tilde F(x,y)$ of \decon\ can be used to
define a function $K_x(\zeta , \bar \zeta)$ on $\IR ^4$ by
$$K_x(\zeta , \bar \zeta)= \tilde F(x^\mu, \zeta_\mu +\bar \zeta _\mu)
\eqn\kis$$
For each $x$, $K_x$ can be viewed as the Kahler potential for a Kahler
metric $G_{\mu \bar \mu}=
\partial _{ \mu} \partial _{ \bar
\mu}K_x$ of signature $(2,2)$
on $\IR ^4$. As a result of \decon, each $K_x$ satisfies the
 equation $det (\partial _{ \mu} \partial _{ \bar
\mu}K_x)=-1$ and
which is often referred to as the Monge-Ampere equation or one of
Plebanski's equations.
The Ricci tensor for the Kahler space is $R_{\nu \bar \nu}=
\partial _{ \nu} \partial _{ \bar
\nu}log \ |det (G_{\mu \bar \mu})|$ and this vanishes if \decon\ holds, so
that
  the   metric is Kahler and
Ricci-flat, which implies that the curvature tensor is either self-dual
or anti-self-dual.
(For Euclidean world-sheets, a similar analysis goes through and leads to a
Kahler Ricci-flat space with Euclidean signature [\hegeom].)

 As the Kahler
potential is independent of the imaginary part of $\zeta _\mu$, the metric
has two commuting (triholomorphic) Killing vectors, given by $i(\partial /
\partial \zeta _\mu - \partial / \partial \bar \zeta _{\bar \mu})$.
Thus the lagrangian $\tilde F(x,y)$ corresponds to a two-parameter family
of Kahler potentials $K_{x^\mu}$ for  self-dual geometries on $\IR ^4$
with two Killing vectors.
The parameter constraint \lamcon\
  implies that $\tilde F +\Lambda$ is  also   a Kahler
potential for a hyperkahler metric with two killing vectors, so that
for each $x$,  $\Lambda$ represents an infinitesimal deformation of the
hyperkahler geometry.
Other relations between $w_\infty$ and self-dual geometry have   been
discussed in [\winf,\park].

Techniques for solving the Monge-Ampere equation can be used to solve
\decon. The general solution of the Monge-Ampere equation  can be given
implicitly
 by   Penrose's twistor transform construction [\pen]. For solutions with one
(triholomorphic) Killing vector, the Penrose transform reduces to a
Legendre transform solution   which was  found first in the context of
supersymmetric non-linear sigma-models [\hit].
Substituting the Legendre transform solution of \decon\
in the action \sis,\fis\
gives precisely the Hamiltonian form of the $w_\infty$ action
(\ie\ the $w_\infty$ generalisation of \soso). In particular, the twistor
space approach of [\hit] gives a twistor interpretation to the auxilairy
fields $\pi$.

For  self-dual spaces with two Killing vectors,
it is possible to write down a new solution of the Monge-Ampere equation,
using a generalisation of the Legendre transform solution that involves
transforming with respect to both components of $y_\mu$.
Any
   $\tilde F(x,y)$ can be written
as a transform of a function $H$ as follows:
 $$\tilde{F}(x
^{\mu},y_{\nu})=2\pi^{\mu}y_{\mu}-{1\over2}\eta^{\mu
\nu}y_{\mu}y_{\nu}-2H(x,\pi)
\eqn\retywt$$
where the equation
$$y_{\mu}={\partial H\over\partial\pi^{\mu}}
\eqn\rtysd$$
implicitly determines
$\pi_{\mu}=\pi_{\mu}(x^{\nu},y_{\rho})$.
Then $\tilde F$ will satisfy \decon\ if and only if its transform $H$
satisfies
$${1 \over 2}
\eta^{\mu\nu}{\partial^{2}H\over\partial\pi_{\mu}\partial\pi_{\nu
}}= {\partial^{2}H\over\partial\pi_{+}\partial\pi_{-}}=1 \eqn\tyghas$$
The general solution of this is
$$H=\pi_{+}\pi_{-}+f(x,\pi_{+})+\bar{f}(x,\pi_{-})\eqn\rtyzzeg$$ This
solution can be used to write the action
$$
S=
\int d^{2}x\left(2\pi^{\mu}y_{\mu}-\eta
_{\mu\nu}\pi^{\mu}\pi^{\nu}-{1\over2}\eta^{\mu\nu}y_{\mu}y_{\nu}-2f(x,\pi
_{+})-2\bar{f}(x,\pi_{-})\right)
\eqn\grhjgdjkgs$$
The field equation for $\pi^\mu $ is \rtysd, and using this to substitute for
$\pi$ gives the action \sis\ subject to the constraint \decon.
Alternatively, expanding the functions $f,\bar f$ as
$
f = \sum  {  s}^{-1}h_s(x)(\pi _+)^s
$, $
\bar f = \sum  {  s}^{-1}\bar h_s(x)(\pi _-)^s
 $
gives precisely the form of the action given in [\pop]. The parameter
constraint \lamcon\ is solved similarly, and the solutions can
be used to write the symmetries of \grhjgdjkgs\ in the form given in [\pop].

\chapter{Quantum \W-Algebras}

Classical \W-algebras were considered in chapter 2.
For those \W-algebras that are Lie algebras, it is usually
straightforward to obtain a corresponding quantum algebra by replacing the
classical currents with quantum field operators, replacing the Poisson
brackets with operator commutators and inserting   factors of $i\hbar$ in
accordance with the minimal Dirac prescription. Then the fact that the
classical algebra satisfies the Jacobi identities usually
implies that the quantum
algebra does also.

For non-linear algebras, however, the situation is more complicated as the
non-linear terms occurring on the right-hand-sides of the commutation relations
involve the products of quantum field operators at the same point and so some
regularisation prescription is necessary. For example, in the classical $W_3$
algebra, the commutator
of two spin-three currents gives rise to the spin-four current
$\Lambda=TT$, whose modes are $\Lambda _n= \sum_m L_{n-m}L_m$, where
$L_n$ are the modes of the stress-tensor $T$.
In the quantum theory, $T(z)T(z)$ is singular and so to
define $\Lambda $ it is necessary to introduce a regularisation.
The following normal ordering prescription is the most convenient:
$$
\op L_n L_m \cl = \cases {L_nL_m  &if $m>n$  \cr
L_mL_n & if $ n \le m$ \cr}
\eqn\norm$$
Using this, the $[W,W]$ commutator   can be written in terms of the
spin-four current with modes
$$\Lambda_{m}=\sum_{n}\op L_{m-n}L_{n} \cl-{3\over10}(m+3)(m+2)L_{m}
\eqn\four$$
(The term  linear in $L_n$ is added to make $ \Lambda_n$
quasi-primary [\zam].)
The regularisation \norm\ corresponds to subtracting
the singular terms in the operator product expansion of $T(z)T(w)$ and
then taking the limit $z \rightarrow w$ and can be generalised
to define the product of any two currents in the \W-algebra.

The full quantum $W_3$ algebra with central charge $c$
[\zam] consists  of the Virasoro algebra
$$
\left\lbrack
L_{m},L_{n}\right\rbrack =
(m-n)L
_{m+n}+{c \over 12} (m^3-m) \delta  _{m+n}.
\eqn\viroo$$
the relation
$$
\left\lbrack
L_{m},W_{n}\right\rbrack =
[2m-n]W_{m+n}
\eqn\primpy$$
so that the spin-three current $W$ (with modes $W_n$) is primary, and
$$\eqalign {
\left\lbrack
W_{m},W_{n}\right\rbrack &=
b^2(m-n)\Lambda
_{m+n}
+{1\over15}(m-n)\left\lbrace(m+n)^{2}-{5\over2}mn-4\right\rbrace L_{m+n}
\cr &+{c\over360}(m^{3}-m)(m^{2}-4)\delta_{m+n}
\cr}\eqn\zamoot$$
where $
b^2 = {16 \over {22+5c}}$. The coefficients are fixed by requiring the
algebra to satisfy the Jacobi identities [\zam].

\section{BRST Charges}

In [\thierry], Thierry-Mieg constructed a BRST charge for the $W_3 $
algebra and this was extended in [\vanbrs]
to a construction of a BRST charge for any quadratically
non-linear \W-algebra; it seems likely
that this can be generalised to any \W-algebra.
For $W_3$, one introduces  the usual conformal ghost system
$b_{++},c_-$ corresponding to $T_{++}$ and a ghost system
$u_{+++},v_{--}$ corresponding to $W_{+++}$.
The BRST charge takes the form
$$
Q=
\int \! \!  d  x^+ \;[c_-(T_{++}- \alpha)+v_{--}(W_{+++}- \beta)+ \dots ]
\eqn\brsch$$
and is nilpotent if
and only if $T,W$ satisfy the $W_3$ algebra with central  charge $c=100$
and the intercepts are $ \alpha =4$ and $\beta =0$ [\thierry].
If the matter system has $c=100$, then this central charge is cancelled by
a contribution of $-26$ from the ghosts $b_{++},c_-$ and a contribution of
$-74$ from the ghosts $u_{+++},v_{--}$.

For a general \W-algebra,
one introduces a spin-$s$ anti-ghost $u^A$ and a spin $1-s$ ghost $v^A$
corresponding to  each spin-$s$ generator $W^A$.
The BRST charge takes the form
$Q \sim \int dx^+( \sum_A v^A W^A + \dots)$ and this can only be nilpotent
if the matter central charge cancels
against the total ghost contribution to the
central charge, although this condition
need not be sufficient in general. The contribution  to the central charge from
the ghosts
for a spin $s$ generator is $c_s= -2(6s^2 -6s +1)$
so that for $W_4$ the
total ghost contribution
is $-246$ while for
 $W_ \infty$ the total ghost contribution
is given by the divergent sum $- \sum_{s=2}^ \infty c_s= -26-74-146- \dots$.
This series can formally be summed using $\zeta$-function regularisation
to give the value $2$ [\zet], but it is not clear  at present what the
significance
 of this result might be.

\chapter {Free Boson Realisations of Quantum Algebras}

In this chapter we will introduce
a number of examples
of free boson realisations  which will be useful
when we discuss \W-gravity anomalies.
In each case we start with a free boson model with
a set of currents that satisfy a Poisson bracket \W-algebra.
We then quantize the boson system and investigate
the quantum algebra generated by the \W-currents.
One might expect the quantum algebra
to be obtained from the classical one
by using the standard Dirac quantisation
prescription of introducing   factors
of $i \hbar$ into the classical relations. In general, however,
the quantum algebra corresponds to the Dirac-quantized algebra plus
extra corrections of order $\hbar ^2$ or higher,
and these corrections can either be central extensions of the algebra
or can involve currents constructed from the boson fields.
Each higher order correction to the algebra corresponds to a
\W-gravity anomaly, as will be seen in the next chapter.

There is a crucial difference between linear and non-linear
realisations of a classical current
 algebra. For linear realisations, the   quantum
 current algebra
 is given by the Dirac-quantized algebra plus central extension terms
 of order $\hbar^2$ and there are no higher order terms.
For non-linear realisations, the situation is altogether more
complicated
as new currents can arise in the quantum algebra which were not
present in the classical
\W-algebra and these reflect a new kind of anomaly which arises for such
models [\wanog].

\section{Linear Realisation of the Virasoro Algebra}

The stress-tensor for $D$ free bosons given by
$$
T= {1 \over 2}   \partial _+ \phi ^i \partial _+ \phi ^i
\eqn\stress$$
(with   $i=1, \dots ,D$) generates the
classical Poisson bracket algebra given by
\con\ with $c=0$, which can be written schematically as
$$[T,T] \sim T
\eqn\vo$$
On quantisation, the modes $\alpha _n^i$ of
$\partial _+ \phi ^i$
are taken to satisfy harmonic oscillator commutation relations
$$[\alpha _n^i,\alpha_m^j]=  \hbar \delta^{ij} \delta_{m+n}
\eqn\harm$$
It is necessary to
regularise \stress, and it is convenient
to define $T= :{1 \over 2}   \partial _+ \phi ^i \partial _+ \phi ^i:$
so that $L_n={1 \over 2}\sum_m   :\alpha_{n+m}^i \alpha_{-m}^i:$,
where colons denote the normal ordering
with respect to the modes $\alpha
 _n^i$:
$$
: \alpha_n ^i \alpha_m ^j : = \cases {\alpha_n^i \alpha_m ^j &if $m>n$  \cr
\alpha_m ^j \alpha _n ^i& if $ n \le m$ \cr}
\eqn\norma$$
Then the commutator  algebra generated by the quantum operator
 $T$ is given by the Virasoro algebra \con\ with central charge $c=D$.
It will be convenient to
suppress delta-functions and numerical factors and present such
algebras schematically as
$$[T,T] \sim  i \hbar T + \hbar^2 c
\eqn\voh$$
The term linear in $\hbar$ is what one would expect from applying the
Dirac prescription to \vo, but there is in addition a central charge term
of order $\hbar^2$, which corresponds to an anomaly, as we shall see.

The stress-tensor \stress\ can be modified by adding a non-minimal
\lq background charge' term, $T \rightarrow T'=T+ a_i \partial _+ ^2 \phi ^i$
where $a_i$ is some constant vector.
The classical Poisson bracket
algebra then has a central charge, $[T',T'] \sim T' + c_0$
where $c_0= {1 \over 24}  a_i a_i$, and the quantum algebra
again takes the form
\voh, but with $c= D + c_0/ \hbar$. This factor of $\hbar ^{-1}$ can
be absorbed into a rescaling of the background charge $a \rightarrow
\sqrt \hbar a$, so that the stress-tensor becomes
$$
T'= {1 \over 2}   \partial _+ \phi ^i \partial _+ \phi ^i  + \sqrt \hbar
a_i \partial _+ ^2 \phi ^i
\eqn\stressbc$$
and the quantum central charge becomes $c=D+a^2/24$.

\section{Linear Realisation of $W_ \infty$}

Let $\phi ^i$ be $D$ complex free bosons. Then the currents [\kir]
$$W_n(x^+)= \sum_{r=1}^{n-1}  \beta_{n,r}   \partial _+^r \phi ^i \partial _+
^{n-r} \bar \phi ^i, \qquad n=2,3,\dots
\eqn\lininf$$
for suitably chosen constants
$ \beta_{n,r}$
 generate a Poisson bracket algebra which is
a certain classical limit of the
 $W_ \infty$ algebra of [\qwinf], (note that this algebra is not the $w_
\infty$ algerba of [\winf])
which has  the generic form [\qwinf, \kir]
$$
[W_n,W_m] \sim W_{n+m-2} + W_{n+m-4} + W_{n+m-6} +   W_{n+m-8} +\dots
\eqn\fhhf$$
This is a linear realisation of a classical limit of $W_ \infty$, as these
currents generate
variations of $ \phi ^i$ which are linear in $ \phi ^i$.
The quantum currents given by normal-ordering \lininf \ generate an algebra
of the general structure (for some constants $c_n$)
$$
[W_n,W_m] \sim
c_n \hbar ^2   \delta_{n,m}
 i \hbar W_{n+m-2} + i \hbar W_{n+m-4} +i \hbar  W_{n+m-6}  + \dots
\eqn\fhhfas$$
which  consists of  the Dirac quantisation of the algebra \fhhf, plus
 central extension terms of order $\hbar ^2$. This quantum algebra contains the
Virasoro algebra with central extension $2D$ and is   the $W_ \infty$
 algebra of [\qwinf] with $c=2D$ [\wanog], generalising the $c=2$
 construction of [\kir].

\section{Non-Linear Realisation of $W_3$}

For $D$ free real bosons, as was seen in chapter 3,
the currents
given by \stress\ and $W= {1 \over 3 } d_{ijk} \partial _+ \phi ^i \partial _+
\phi ^j \partial _+ \phi ^k$ generate a classical $W_3$ algebra provided
the constants $d_{ijk}$ satisfy \did. The corresponding transformations of
$\phi ^i$   are given by \sym\ and are non-linear for spin 3.
The classical algebra \con,\prim,\walg\ can be written schematicaly as \vo\
together with
$$[T,W] \sim W, \qquad [W,W] \sim \Lambda, \qquad \Lambda \equiv TT
\eqn\wth$$

In the quantum theory, the currents can be defined using the normal
ordering \norma.
The spin-two currents again generate the Virasoro algebra \voh\ with
$c=D$. The $[T,W]$ commutator now takes the form
$$ [T,W] \sim i\hbar W + \hbar^2 J  , \qquad J \equiv {d^i}_{ij} \partial _+
\phi ^j
\eqn\qwerty$$
consisting of the Dirac prescription term, plus a term of order $\hbar^2$
which involves a new spin one current, $J$ [\he].
Thus although the classical algebra closes, the quantum one does not
unless the $d$-tensor is traceless,
${d^i}_{ij}=0$.
Even if it is traceless, the $[W,W]$ commutator
gives
$$
[W,W] \sim i \hbar :TT: + (D+2) \hbar^2 (T^{2,0} +T)+ \hbar^3 c',
\qquad T^{2,0} \equiv :\partial _+ \phi ^i \partial _+ ^3 \phi ^i :
\eqn\qwth$$
which consists of the Dirac term, a central charge term
of order $\hbar^3$ proportional to $c' \equiv d_{ijk}d^{ijk}$
and a new spin-four current
$T^{2,0}$ [\he,\wama]. If ${d^i}_{ij} \ne 0$ there are extra
terms in \qwth\ involving
$J$ [\wama]. Since the coefficient of the spin-four current
$T^{2,0}$ is non-zero for any $D>0$, the quantum algebra never closes
when the normal
ordering prescription \norma\ is used, as the right hand side of
\qwth\ cannot be written entirely in terms of $T,W$ and
composites constructed using \norma, such as $:TT:, :TW:$ etc.
To close this algebra, it is necessary to introduce
$J,T^{2,0}$ as generators, and then to introduce further generators, such as
$T^{n,0}= \partial _+^{n+1} \phi ^i \partial _+ \phi ^i, W^{m,n,0}=
 d_{ijk} \partial _+ ^m \phi ^i \partial _+ ^n \phi ^j \partial _+ \phi ^k,
\dots$ [\wama].

However, instead of defining the composite operator $:TT:$ using
the prescrition \norma, it could instead be defined as $ \op TT \cl $
using \norm. The two definitions are related by [\he]
$$\op TT \cl =:TT: - \hbar T^{2,0}
\eqn\normdif$$
so that they differ by a finite term of order $\hbar$.
Using this, the algebra \qwth\ can be rewritten as
$$
[W,W] \sim i \hbar \op TT \cl + (D-2) \hbar^2 (T^{2,0}+T) +\hbar ^3 c',
\qquad T^{2,0} \equiv \partial _+ \phi ^i \partial _+ ^3 \phi ^i
\eqn\qwthu$$
so that the coefficient of the   spin-four current $T^{2,0}$ is now
$D-2$ instead of $D+2$, with the result that the algebra closes
non-linearly on $T,W, \op TT \cl$ if and only if
${d^i}_{ij}=0$ and $D=2$ [\he]. For $D=2$, the solution of \did\ given by
$d_{112}=-\kappa $ and $d_{222}=\kappa$ gives a traceless $d$-tensor
and hence a closed algebra which becomes preciesly
the $W_3$ algebra \viroo,\primpy,\zamoot\
after rescaling the currents. This is the two boson realisation of the
$c=2$ $W_3$ algebra given in [\fat].

This model is closely related to the
Casimir construction of $W_3$ [\thierrya,\bais].
 The classical $W_3$ algebra is
realised by the Sugawara
currents $T= {1 \over 2} tr (J_+J_+)$ and
$W= {1 \over 3} tr (J_+J_+J_+)$,
 where $J_+$ is an $SU(3)$ Kac-Moody
current [\heee]. In the quantum theory, these
currents (after normal ordering)
no longer generate a closed algebra, as the commutator $[W,W]$
gives rise to
 the spin-four current $T^{2,0}=
 tr(J_+ \partial _+ ^2 J_+ )$ [\thierrya,\bais],
 which is similar to the current $T^{2,0}$ that arose in the free boson model.
 In the case in which the Kac-Moody algebra is of level one, it
 is possible to perform a truncation to a realisation of the
  quantum $W_3$ algebra, and this is related to the fact that
  the level one Kac-Moody algebra can be constructed from the two
  boson model discussed above [\bais].

\section{$W_3$ Realisations with Background Charges}

As in the Virasoro case, one can consider adding higher derivative
terms to the currents to give
$$
T'=T+ \sqrt \hbar a_i \dpl ^2 \ffi, \qquad W' =W + \sqrt \hbar e_{ij} \dpl
^2\ffi \dpl \fj + \hbar f_i \dpl ^3 \ffi
\eqn\curmod$$
for some constants $a_i,e_{ij},f_i$.
The current
 algebra will close
on $T,W, \op TT \cl $ to give the $W_3$ algebra \viroo,\primpy,\zamoot\
provided the tensors $a_i,e_{ij},f_i$ satisfy certain
constraints [\he]. For the $D=2$ model these constraints are satisfied by
choosing
the only non-vanishing components to be given in terms of a free parameter
$a$ by $a_1=a,e_{12}=a,e_{21}=3a, f_2=6a^2$, giving a
model with central charge  $c=2+24a^2$ [\fat]. For real $a$ one obtains models
with any value of $c \ge2$, while for imaginary $a$,
a unitary theory can only be defined
if the   background charge $a$ is chosen to take the
discrete values $a^2=-[p(p+1)]^{-1}$ for $p=4,5,6,\dots$ giving the minimal
series of representations  of the {\cal W}-algebra with central charge
 $c_p=2-24[p(p+1)]^{-1}$ [\fat].
 For $D \ne 2$, the constraints on the coefficients in \curmod\ were solved
 in [\romans], giving a realisation of the $W_3$ algebra in terms of any
 number $D$ of bosons, with arbitrary central charge
$c=D+a^2/24$. In particular, it is possible to construct in this way
realisations of $W_3$ with central charge $c=100$ for which
there is a nilpotent BRST operator.

 One can instead ask whether modifying the currents as in \curmod\ can
 give an algebra that closes using the normal ordering \norma, \ie\ whether
 an algebra can be found  in which $[W,W]$ can be written
entirely in terms of
  $T,W$ and $:TT:$
  instead of $T,W$ and $\op TT \cl$.
 It was shown in [\pal] that this is only possible if there are
 precisely $D=2$ bosons,
 in which case $a_i=f_i=0$ and $e_{ij}$ is proportional to
 $ \epsilon _{ij}$, so that the Virasoro subalgebra has $c=2$.

\section{One Boson Realisation of $W_ \infty$}

For one boson $\phi$, the currents
$W_n= {1 \over n} ( \partial _+ \phi)^n$ for $n=2,3, \dots$
generate the $w_ \infty$ algebra classically [\pop]. The
$w _ \infty$ algebra    takes the schematic form
$$
[W_n, W_m] \sim W_{n+m-2}
\eqn\ruuo$$
All the currents except the Virasoro current $W_2$ generate
non-linear transformations and this non-linearity leads to new
currents on the right-hand-sides of the quantum
commutation relations e.g.
$[W_2,W_3]$ gives rise to the current $J= \partial _+ \phi$ while
$[W_3,W_3]$ gives rise to $T^{2,0}= :\partial _+ \phi \partial _+ ^3 \phi
:$.
However, one can again consider adding higher derivative terms to the currents
$W_n$ (as in \curmod) to modify the algebra. It was shown in [\deform]
that the coefficients can be chosen in such a way as to close the algebra,
giving rise to the $W_ \infty$ algebra of [\qwinf],
which is of the schematic form \fhhf,
 with central charge
$c=-2$. This is precisely what is needed to cancel the contribution
of $+2$ coming from using a $\zeta$-function to sum the infinite
number of ghost contributions [\deform].

\chapter{\W-Gravity Anomalies from Matter Integration}

There has been a great deal of recent work on the quantisation
of \W-gravity and the
anomalies that arise [\wanog -\wstr].
The classical coupling of a free boson system to \W-gravity is
described by an action of the form \newtq, where $S_0$ is the free boson action
\free\ and the $h^{ \pm A}$ are gauge fields.
In this chapter we will consider the
integration over the matter fields $ \phi ^i$ only, regarding the
gauge fields as external sources, and study the anomalies that arise in the
Ward identities corresponding to
the classical \W-gravity symmetries. No gauge fixing is needed as the
gauge fields are not being integrated over, and the scalar fields
have a well-defined propagator.
In the next chapter, we will gauge fix all the local symmetries,  introduce the
appropriate ghost fields and discuss the integration over all fields.

For simplicity, consider a chiral \W-gravity, so that
the complete action is of the form \newt, consisting of the free boson action
\free\ plus the linear Lagrange multiplier terms.
Since this is just a free system plus constraints,
the normal ordering prescription \norma\ is sufficient
to subtract all divergences. It is convenient to use
the background field method, writing $\phi ^i$ as the sum $\bar \phi ^i +
\varphi^i$
of a background field $\bar \phi ^i$ and a quantum field $\varphi^i$
and then integrating over $\varphi^i$ to obtain a
(renormalised) effective action
$ \Gamma[ \bar \phi ^i, h^{+A}]$, which is given by the classical action \newt,
plus quantum corrections $ \Gamma[ \bar \phi ^i, h^{+A}]= S
[ \bar \phi ^i, h^{+A}]+O( \hbar)$.
The classical action is invariant
under the classical \W-symmetries and these lead to Ward identities
for $\Gamma$.
These are of the form
$
\delta \Gamma + \dots = \Delta \cdot \Gamma$
where $\delta \Gamma$ is the gauge variation of the effective action,
$\Delta$ is a local operator
and
$\Delta \cdot \Gamma$
denotes
all 1PI graphs with precisely one   insertion  of  $\Delta$.
The dots denote other terms in the Ward identity, the details of which are
discussed in [\brink,\baswar] and which will not be important here.
For non-anomalous theories, $ \Delta$ is zero or
can be cancelled by adding finite local counterterms to the action.

We will now present the anomalies for the models given by
gauging the bosonic realisations of \W-algebras described in the
previous chapter.

\section{Gravitational Anomaly}

Consider first the free boson model coupled to chiral gravity
 with action
$S=S_0-
\int \! \!  d^2 x \; h_{--}T_{++}$ with $T_{++}$ given by \stress.
The effective action takes the form
$$
\Gamma=S+ {i  \over 2 \hbar}
\int \! \!  d^2 x d^2 y\; h_{--}(x) h_{--}(y)<T_{++}(x)
T_{++}(y)>+ \dots
\eqn\effo$$
and varying this under a chiral diffeomorphism with parameter $k_-$
leads to the standard chiral gravitational anomaly (to linearised  order
in the gauge field $ h_{--}$) [\pol]
$$
\Delta_{gravity} = D {i \hbar \over 24 \pi}\int \! \!  d^2 x \;k_- \partial
_+^3  h_{--}
\eqn\gravan$$
This anomaly, which is proportional to $D$, corresponds to the central charge
term in \voh, which is also proportional to $D$. The relation between
the two
can be seen by noting that the central term in the commutator
$[T,T]$ leads to a term proportional to $D p_+^4 /p^2$ in the momentum space
correlation function $<T_{++}(p)T_{++}(-p)>$. Substituting this in \effo\
gives
$$\Gamma
 = S_0 +  D {i \hbar \over 24 \pi}\int \! \!  d^2 x \;
 h_{--} {\partial _+^4 \over \boxx} h_{--}
 + \dots
 \eqn\effact$$
 and varying this using $ \delta h_{--} = \partial _-k_-+ \dots$
 leads to the anomaly \gravan.
If the stress-tensor is modified to become \stressbc, then the anomaly is again
given by \gravan, but with $D$ replaced by $D+ a^2/24$.

\section{Linear $W_ \infty$ Gravity}

Gauging the linear realisation of chiral $W_ \infty$ [\kir] reviewed
 in section 8.2
gives a theory with spin-$n$ gauge fields $h_n$ and symmetries
with parameters $ \lambda_n$ of spin $n-1$, with transformations of the form
$\delta h_n = \partial _- \lambda_n +\dots$, for $n=2,3,\dots$.
The anomaly takes the form [\wanog]
$$
\Delta= \sum_n c_n \hbar \int \! \!  d^2 x \; \lambda_n \partial _+ ^{2n-1} h_n
\eqn\linwinfanom$$
for some coefficients $c_n$
and these terms  correspond precisely to
the $O( \hbar^2)$ central extension terms  in \fhhfas.
The $n=2$ term is the gravitational anomaly \gravan.

In this and the previous example, the   anomalies   depend only
on the gauge fields and not on the background matter fields $\bar \phi ^i$
and correspond to central charge terms in the current algebra. All
anomalies that
occur in linearly realised symmetries are of this type and we shall
refer to such anomalies as {\it universal anomalies} [\wanog].

\section{Anomalies of Chiral $W_3$ Gravity}

Chiral $W_3$-gravity has an action
given by the sum of \free\ and
\noeth,
 with gauge fields $h_{--}, B_{---}$.
The linearised anomaly for the $k_-$ and $\lambda_{--}$ symmetries
is the sum of the following terms [\he,\hee,\wama]: the
gravitational anomaly \gravan, the mixed
spin-2/spin-3 anomaly
$$ \Delta_{mixed}=
-{i \hbar \over 12 \pi} \int \! \!  d^2 x \;
\bar J_+ (B_{---} \partial _+^3 k_- + \lambda_{--}
\partial _+ ^3 h_{--}),
\eqn\mixed$$
 the universal spin-3 anomaly [\mato]
$$
\Delta_{\W-univ}
=- c' { \hbar^2 \over 1440 \pi^2} \int \! \!  d^2 x \; \lambda_{--} \partial
_+^5 B_{---},
\eqn\univ$$
and finally the remainder of the
spin-3 anomaly
$$\Delta_{ mat }=
- {i \hbar (D+2)\over 24 \pi}
\int \!    d^2 x \; \Bigl[ 3  \bar T^{2,0}( \lambda \partial _+ B- B \partial
_+
\lambda)+\bar T (-3 \partial _+ \lambda \partial _+^2 B + \dots)+ O( \bar J)
\Bigr],
\eqn\matter$$
Here
$$
\bar J \equiv {d^i}_{ij} \partial _+
\bar \phi ^j,
 \qquad \bar T^{2,0} \equiv \partial _+ \bar \phi ^i \partial _+ ^3 \bar \phi
^i
, \qquad \bar T= {1 \over 2} \partial _+ \bar \phi ^i \partial _+ \bar
\phi ^i
\eqn\eryuytyu$$
are the currents $J,T ,T^{2,0}$ of section 8.3 for the background fields $\phi
^i$.
These results were confirmed and extended to higher orders in the
gauge fields in [\vanan].

The universal spin-3 anomaly \univ\ corresponds to the central term in \qwth\
but arises at two loops instead of one-loop, as a result of the
non-linearity of the symmetry.
The mixed anomaly \mixed\ depends on the scalar fields through the current $
\bar  J$ and corresponds to the $J$-dependent term in \qwerty.
Similarly, the anomaly \matter\ depends on the
scalar fields through the currents $ \bar T , \bar T^{2,0} $
 corresponding to the
$T ,T^{2,0}$ terms in \qwth\ which are also proportional to $D+2$, and through
the current $\bar J$ corresponding to the $O(J)$ terms which were suppressed in
\qwth.

It should be stressed that the theory is regularised using the
normal ordering prescription \norma, and that the prescription \norm\
would not be sufficient to remove all the
divergences from the path integral. For this reason, the anomaly
corresponds to the algebra \qwth\ written using the normal ordering
\norma\ and so has a coefficient $D+2$ as in \qwth, instead of the $D-2$ that
occurs in the algebra \qwthu\ that uses \norm.

For obvious reasons, the  non-universal anomalies depending on the
matter currents will
be referred to as {\it matter-dependent anomalies} [\wanog].
 They do not arise for linearly realised symmetries but usually occur for
non-linearly realised
ones.
The part of the anomaly \matter\ which is proportional to
$\bar T$ can be cancelled by modifying the $ \lambda$-transformation of the
field $ h_{--}$ by a term proportional to
$ \hbar (D+2)(-3 \partial _+ \lambda \partial _+^2 B + \dots)$
but the terms proportional to $ \bar T^{2,0}$ and $\bar J$ appear to be
non-trivial anomalies.

To check whether the anomalies are non-trivial,
it is necessary to investigate whether they can be cancelled by
adding finite local counterterms.
Consider adding   counterterms of the form [\hegho, \wstr]
$$S_{count}
=
 -\int \! \!  d^2 x \; \Bigl[
 \sqrt \hbar  a_i \dpl ^2 \ffi  h_{--}
 + (\sqrt \hbar e_{ij} \dpl
^2\ffi \dpl \fj + \hbar f_i \dpl ^3 \ffi)B_{---} \Bigr]
\eqn\actmod$$
to the $W_3$ gravity action
so that the gauge field term is now proportional to
$ hT' + BW'$ where $T',W'$ are the modified currents in \curmod.
These appear to be the most general terms that can be added without introducing
new fields or dimensionful couplings and which might affect the linearised
anomaly.
There are   values of the coefficients
$a_i,e_{ij},f_i$ that lead to a cancellation of all the matter-dependent
anomalies (after appropriate
modifications of the transformation rules)
 if and only if the number of bosons is $D=2$, in which case
$a_i=f_i=0 $ and $ e_{ij}$ is proportional to
$ \epsilon_{ij}$. These were precisely the values that led to the
current algebra which closes on $T',W', :T'T':$ [\pal].
Thus the conditions that the theory obtained by gauging
the algebra generated by $T',W'$ be free of matter-dependent anomalies
are precisely the conditions that $T',W'$ generate a closed \W-algebra
(with the normal ordering prescription \norma).
However, even in this case, there remain non-trivial universal anomalies.
For $D \ne 2$, the matter-dependent anomalies are non-trivial.

\section{Non-Linear  $w_ \infty$ Gravity}

Chiral $w_ \infty$ gravity is given by introducing the Noether coupling
$ \sum_n  h^n W_n$ of the gauge fields $h^n$ to
the currents \wiss\ [\pop].
Integrating out the scalar  field $ \phi$ gives contributions to the effective
action of the form
$$\sum _{m,n}a_{m,n}
\int \! \! d^2x d^2y \; h^m(x) h^n(y)<W_m(x) W_n(y)>$$
(for some coefficients $a_{m,n}$)
and varying this under $ \delta h^n = \partial _- \lambda^n + \dots$
gives an infinite  set of anomalies similar to those described
for $W_3$.
There are matter-dependent anomalies for all spins higher than two, as
all the higher spin symmetries are non-linearly realised.
It was argued  in [\deform] that all the matter-dependent
anomalies can be cancelled by adding higher derivative counterterms
similar to \actmod. The action with the counterterms
now involves the coupling
$ \sum_n  h^n W'_n$ where $W'_n$ are modified currents which are precisely the
currents
that generate the quantum algebra $W_ \infty$ with $c=-2$ [\deform].
There remain non-trivial universal anomalies (one for each spin),
including a gravitational anomaly proportional to $c=-2$.
These can cancel against ghost contributions, however,
as will be seen in the next chapter.
Unfortunately, this cancellation of anomalies does not seem to be possible
for the realisations of $w_ \infty$ in terms of
more than one boson [\hegho].

\section{ Anomalies and Current Algebras}

The previous examples are sufficient to illustrate the general principle
that non-trivial  anomalies arise if the corresponding
 current algebra closes classically but not quantum
 mechanically.
In general the quantum current algebra is given to lowest order
in $ \hbar$ by applying the
Dirac prescription to the classical algebra. However, there will
in general be extra  terms of order $\hbar ^2 $ or higher and each of these
terms
corresponds to  a term in the anomaly.

The matter-dependent terms in the current algebra
of order $\hbar ^2 $ or higher fall into two classes,
depending on whether they involve only the currents $W_A$ that
occurred
in the
classical algebra (e.g. $ \{ W_A \} = \{ T,W  \}$ for
$W_3$), or whether they introduce new currents into the algebra
(e.g. $J,T^{2,0}$ for
$W_3$).
Those which involve only the currents $W_A$ do not affect the closure
of the algebra but represent an $\hbar$-dependent  deformation
of the classical \W-algebra.
The
corresponding terms in the anomaly
depend on the currents $W_A$.
The classical action \newt\ involves the term $ \sum _A h^A W_A$ and so
   these $W_A$-dependent anomalies can be  cancelled by modifying the
 gauge  transformations of the gauge fields $h^A$ by
 $ \hbar$-dependent terms.
 An example of this kind of term is the term proportional to $T$ in \qwth,
 which leads to the term proportional to $ \bar T$ in the anomaly \matter;
 this term
  was cancelled by modifying the transformation of $h_{--}$.

 Matter-dependent anomalies of this type were first found
 in [\gates] in models with a non-linearly realised local
 supersymmetry.
 For these models, there were universal anomalies as usual, together with
 a matter-dependent supersymmetry
 anomaly involving the stress-tensor $T_{++}$.
  The matter-dependent
 anomaly could then be cancelled by modifying the supersymmetry transformation
 rule for the graviton $ h_{--}$ by appropriate $
 \hbar$-dependent terms.
 This  corresponded to a deformation of the classical symmetry algebra,
 which was of the form $Q^2=0$, where $Q$ was the classical \lq supercharge',
 to a quantum (1,0) supersymmetry algebra
  of the form $Q^2= \hbar P$, with $P$ the chiral momentum [\gates].

 If there are matter-dependent terms in the current algebra
 that involve new currents not in the classical algebra, then
 the quantum algebra does not close on
 the set of classical currents $ \{ W_A \}$. For example,
 the presence of
 $T^{2,0}$ in \qwth\
 means that the algebra  does not close on the classical currents $T,W$.
 There are corresponding
 terms in the anomaly involving the  new currents (e.g. the $ \bar T^{2,0}$
 term in \matter) and these cannot be cancelled by modifying the transformation
 rules.

 One way of cancelling such matter-dependent anomalies is as follows
 [\wama]. The set of currents $W_A$ generate an algebra which closes
 classically but not quantum mechanically.
 One can introduce a (possibly infinite) set of
 new currents into the algebra until one obtains a quantum algebra
 (with generators $Z^m$, say) which
 is closed up to central charge terms.
For example, in the case of $W_3$, one introduces $T^{2,0}$ as a generator,
 and then finds
 that the algebra generated by $T,W,T^{2,0}$ still does not close, and
 so one is led to introduce an infinite set of currents
 $T^{n,0}= \partial _+^{n+1} \phi ^i \partial _+ \phi ^i, W^{m,n,0}=
 d_{ijk} \partial _+ ^m \phi ^i \partial _+ ^n \phi ^j \partial _+ \phi ^k,
\dots$ which do generate a closed quantum algebra.
 If instead of gauging the original \W-algebra, one gauges the
 its quantum closure by introducing a gauge field
 corresponding to each of the generators $Z^m$, then there will
 be no non-trivial matter-dependent anomalies and
 only the universal anomalies will remain [\wama].

Central charge terms in the algebra correspond
to universal anomalies and these are non-trivial
(\ie\ anomalies which cannot be cancelled by local counterterms)
if and only if the central extension is a non-trivial cocycle (\ie\ one
which cannot be absorbed into redefinitions of the generators).
If the classical algebra had no central charges, the quantum generation
of such terms means
that the quantum algebra does not close on the original set of generators
but requires the addition of a central charge generator. In this case, one can
follow the approach outlined above and
introduce a spin-zero gauge field which couples to the generator
of central charge gauge transformations [\heeee], so that one is gauging
an algebra which closes properly at the quantum level, not just up to
central charge terms. However, it is more conventional
not to do this, but instead to try and cancel the central charge terms
against ghost
contributions.

\section{Quantum \W-Gravity}

The effective action for chiral gravity was given to lowest order by
\effo. For the non-chiral case, it is given instead by
 $$\Gamma
 = S_0 +  D {i \hbar \over 24 \pi}\int \! \!  d^2 x \;
 \left[
 h_{--} {\partial _+^4 \over \boxx} h_{--}
 +
 h_{++} {\partial _-^4 \over \boxx} h_{++}
 + \dots \right]
 \eqn\effactnc$$
 to lowest order in the gauge fields. This is not gauge invariant,
 but can be made so by introducing the trace of the metric, $h_{+-}$, which
transforms as $ \delta h_{+-}= \partial _+ k_- + \partial _- k_++ \dots$, and
 adding a finite local
 counterterm to \effactnc, so that it becomes [\pol]
 $$\Gamma
 = S_0 +  D {i \hbar \over 24 \pi}\int \! \!  d^2 x \;
 \left[
 R^{(2)} {1\over \boxx} R^{(2)}   \right]
 \eqn\effactncr$$
where $R^{(2)}$ is the curvature scalar.
This is gravitationally covariant, but not Weyl-invariant since
it depends on the trace of the metric. Thus the gravitational anomaly has
been cancelled at the expense of
introducing a Weyl anomaly.
Note that the integration over ghost fields shifts the coefficient $D$ in
these formulae to $D-26$.

If $D \ne 26$, the action $R^{(2)} {1\over \boxx} R^{(2)} $
can be thought of as a non-local
kinetic term for the two-dimensional metric, which becomes a
dynamical field. In the conformal gauge, the metric is given in terms of
the  conformal mode $ \phi ^{(2)}= h_{+-}$ and the action takes the local
form $ \phi ^{(2)} \boxx \phi ^{(2)}$ [\pol]. In the chiral  gauge,
the action takes the non-local form $ h_{--} \partial _+^4 \boxx ^{-1} h_{--}$
[\chirgag].

For $W_3$-gravity, the universal spin-3 anomaly gives an extra
contribution to the effective action, which is given to lowest order by
$$
\int \! \!  d^2 x \;
\left[
 B_{---} {\partial _+^6 \over \boxx} B_{---}
 +
 B_{+++} {\partial _-^6 \over \boxx}B_{+++}
 + \dots \right]
 \eqn\effactbb$$
 In this case, there are several ways to introduce new variables and add
 finite local counterterms to cancel the \W-gravity
 anomaly at the expense of introducing a \W-Weyl anomaly [\wanog,\frau].
 As in chapter 6, one can introduce  the traces $B_{ \pm +-}$
 for the spin-3 field $B_{ \mu \nu \rho}$, which
 can then be used to define the invariant curvature given to linearised
 order by [\wanog]
 $$R^{(3)}= \epsilon ^{ \mu \alpha} \epsilon ^{ \nu \beta}
 \epsilon ^{ \rho \gamma}  \partial _\mu \partial _\nu \partial _\rho
 B_{ \alpha \beta \gamma}
 \eqn\ris$$
Then adding a finite local counterterm brings the
action \effactbb\ to the form
$R^{(3)} {1\over \boxx} R^{(3)}$ which is invariant under
the \W-transformation $ \delta B_{ \alpha \beta \gamma}
 = \partial_{( \alpha} \lambda_{ \beta \gamma)}+ \dots$
 but is anomalous under the \W-Weyl transformations
 corresponding to   shifts of the traces,
   $ \delta B_{ \pm +-}= \Omega_ \pm$ .
In a generalisation of the conformal gauge, the traces
$B_{ \pm +-}$ are written in terms of a scalar $ \phi^{(3)}$ and the
$W_3$-gravity action takes the higher-derivative form
$ \phi ^{(2)} \boxx \phi ^{(2)}$ $+\phi ^{(3)} \boxx ^2\phi ^{(3)}
$[\wanog]. In a chiral gauge  the action takes the form
$B_{---} \partial _+^6
\boxx ^{-1} B_{---}$ to lowest order in the gauge fields [\wanog].
Further details, including
  non-linear corrections in the chiral
gauge, are discussed in [\vanan,\vanangha].

Instead of introducing the vector
$B_{ \mu +-}$, one can introduce a
scalar $ \chi$ and define an invariant curvature
$ \bar R^{(3)}= \partial _+ ^3 B_{---}+ \partial _-^3 B_{+++} + \boxx \chi$
[\wanog]. This can be used
to add finite local counterterms so that the action again takes the
 covariant form $ \bar R^{(3)} {1\over \boxx} \bar R^{(3)}$.
  This is not invariant under the \W-Weyl transformation
corresponding to shifts in $ \chi$.
In this case, the linearised conformal gauge
$W_3$ action takes the form
$ \phi ^{(2)} \boxx \phi ^{(2)}$ $+\chi \boxx   \chi
$ which does not involve any higher derivative terms.
Thus apparently inequivalent versions of  quantum \W-gravity can
be obtained by using different ways of introducing conformal
modes into the covariant theory [\wanog].
For further discussion of quantum \W-gravity, see
[\awa,\mat,\vanan,\vanangha].

\chapter {Quantum \W-Gravity and \W-Strings}

Consider first chiral $W_3$ gravity.
 The gauge symmetries \vary\ can be used to gauge
away the gauge fields and impose the gauge conditions $ h_{--}=B_{---}=0$,
or to impose a background gauge
$h_{--}= \bar h_{--}$, $B_{---}= \bar B_{---}$ for some fixed background
gauge fields $\bar h_{--}, \bar B_{---}$.\foot{Strictly speaking,
the gauge fields cannot be gauged away completely in general,
but can be set equal to quadratic and cubic differentials respectively.
The integration over the gauge fields then reduces to an integration over
these spaces of differentials, which constitute the moduli space  for
$W_3$ gravity.}
In doing so, it is necessary to introduce the
usual gravitational conformal ghosts
$b_{++},c_-$ together with their spin-three counterparts $u_{+++},v_{--}$.
The Faddeev-Popov method cannot be used as the gauge algebra does not close
off-shell, but the more general methods of Batalin and Vilkovisky
[\batalina]
can, and yield the following
gauge-fixed action [\he]
$$S =S_0- {1\over2}\int \! \!  d^2 x \;
 \left[ \bar h_{--}(T+T^{gh})+
 \bar B_{---}(W+W^{gh})\right]
\eqn\ghnoeth$$
where the ghost spin two and three currents are
$$\eqalign
{T^{gh}_{++}&=b_{++}\partial_{+}c_{-}+\partial_{+}\mathop{(b}\nolimits
_{++}c_{-})+u_{+++}\partial_{+}v_{--}+\mathop{2\partial}\nolimits
_{+}\mathop{(u}\nolimits_{+++}v_{--})
\cr
W^{gh}_{+++}&=2u_{+++}\partial_{+}c_{-}+\partial_{+}\mathop{(u}\nolimits
_{+++}c_{-})+\mathop{(b}\nolimits_{++}\partial_{+}v_{--})T_{++}+\partial
_{+}(b_{++}v_{--}T_{++})
\cr}\eqn\ghcur$$
Integration over the ghost fields as well as the
matter fields gives further contributions to the anomalies
of section 9.3, and also gives new
matter-dependent anomalies which depend on the ghosts instead of the matter
fields [\vanangh,\hegho]. The ghosts change the coefficient of the
gravitational anomaly \gravan\
 from $D$ to $D-100$ and change the coefficient of the
$T^{2,0}$ term in \matter\
from $D+2$ to $D-2$ [\vanangh].
This latter change is particularly striking, as it is similar to the
change from the algebra \qwth\ to the algebra \qwthu.

The next step is to attempt to cancel the anomalies
by adding finite local counterterms.
The current $W^{gh}$ is primary with respect to $T^{gh}$ but is only primary
with respect to the total energy-momentum $T+T^{gh}$ if it
is modified to give [\he]
$$
\hat W^{gh} =W^{gh} + {\hbar c_m \over 192}
[-2v \dpl ^3 b -9 \dpl v \dpl ^2 b +15 \dpl b \dpl ^2 v + 10b \dpl^3v].
\eqn\wmod$$
The relation between current algebras and anomalies described in the
last section suggests replacing $W^{gh}$ with $\hat W^{gh}$ in
\ghnoeth\ in order to avoid mixed anomalies.
The analysis of the last chapter   suggests adding background charges so that
the matter currents $T,W$  are replaced by the currents \curmod.
It then remains to see whether the free coefficients can be chosen so
that the anomalies cancel [\hegho].
Remarkably, all the anomalies are cancelled by ghost contributions
 and the effective action satisfies the BRST Ward-identities if
 and only if the currents $T',W'$ are chosen so that they generate
 Zamolodchikov's $W_3$ algebra with $c=100$ [\wstr] and for these theories
 the nilpotent
 BRST charge is precisely the $W_3$ BRST charge given in [\thierry].
 This analysis can be extended to non-chiral $W_3$  gravity,
 with the result that the anomalies cancel if
  the currents $T_{++},W_{+++}$ and $T_{--},W_{---}$
  in the action \noether\
 generate two copies of the  $c=100$ $W_3$ algebra [\wstr].
 Thus the $c=100$ realisations of  $W_3$  that were constructed in
 [\romans] can be used to define critical $W_3$  strings for which
 all the anomalies cancel and the integration over gauge fields
 reduces to an integration over moduli.

It seems likely that similar critical strings can be defined
for other \W-algebras.
For the linearly realised $W_ \infty$ gravity of section
9.2, the quantum gauge algebra is $W_ \infty$ with $c=2D$, while for
the non-linearly
realised  $W_ \infty$ gravity of section 9.4, the quantum gauge algebra is
  $W_ \infty$  with $c=-2$ after the cancellation of the matter-dependent
  anomalies. In both cases, the integration over the ghost fields does
  not introduce any new matter-dependent anomalies, so the theories
  will be
  anomaly free if and
  only if the ghost contributions to the
  universal anomalies cancel the matter contributions.  The ghost contributions
to the
  anomalies are given by a divergent sum, which can be regularised
  and summed
 using the zeta-function technique [\zet]
 and the anomalies will then
 cancel only if the matter central charge
 is $c=-2$.
 Thus for the linear realisation, the anomalies do not cancel for any
 value of the number of bosons $D$, but for the one-boson
 non-linear realisation, all the anomalies cancel [\deform] (if the
zeta-function trick is used)
 and a critical $W_ \infty$ string appears to emerge.

  Much work remains to be done to see whether these
 \W-string theories can be extended to fully interacting models with
 a sensible space-time interpretation; some preliminary discussion of the
 spectrum is given in
 [\wstrbil,\das,\wstrspec]. Nevertheless,
 it is indeed a striking result that the standard string theory, which is
 based on the Virasoro algebra, has a generalisation to
 a critical \W-string theory based
 on the $W_3$ algebra. It seems reasonable to expect
 that these results will extend to other \W-algebras,
 giving an infinite  set of new string theories.
Whether or not they will have important physical implications is not yet clear,
but it seems likely that the mathematical
structure of these theories will continue to fascinate researchers for some
time to come.

\refout
\bye